%% file: NFFL2021.tex
\documentclass{article}

\usepackage[final]{NFFL2021}

\usepackage[utf8]{inputenc} 
\usepackage[T1]{fontenc}    
\usepackage{hyperref}       
\usepackage{url}            
\usepackage{booktabs}       
\usepackage{amsfonts}       
\usepackage{nicefrac}       
\usepackage{microtype}      
\usepackage{xcolor}         
\usepackage{amsmath}
\usepackage{mathtools}
\usepackage{amssymb,amsfonts,amsthm}
\usepackage{wrapfig}
\usepackage{acronym}
\usepackage{comment}
\usepackage{hyperref}
\usepackage{url}
\usepackage{bm}
\usepackage{hyperref}
\usepackage{amsthm}
\usepackage{amsmath}
\usepackage{esvect}
\usepackage[noend]{algpseudocode}
\usepackage{centernot}
\usepackage{algorithm}
\usepackage{subcaption}
\usepackage{bbold}
\usepackage[mathscr]{euscript}

\newcommand{\nbb}{\mathbb{N}}
\newcommand{\rbb}{\mathbb{R}}
\newcommand{\ubb}{\mathbb{U}}

\newcommand{\norm}[1]{\left\lVert#1\right\rVert}
\newtheorem{theorem}{Theorem}

\newtheorem{proposition}[theorem]{Proposition}
\newcommand{\E}{\mathbb{E}}
\title{Minimal Model Structure Analysis for Input Reconstruction in Federated Learning}

\author{%
  Jia Qian,  Hiba Nassar, Lars kai Hansen \\
  Department of Applied Mathematics and Computer Science,\\
  Technical University of Denmark,\\
  2800 Lyngby, Denmark.\\
  \texttt{\{jiaq,hibna,lkai\}@dtu.dk} \\
}

\begin{document}
\input{00Acron}
\maketitle

\begin{abstract}
\ac{fl} proposed a distributed \ac{ml} framework where every distributed worker owns a complete copy of global model and their own data. The training is occurred locally, which assures no direct transmission of training data. However, the recent work \citep{zhu2019deep} demonstrated that input data from a neural network may be reconstructed only using knowledge of gradients of that network, which completely breached the promise of \ac{fl} and sabotaged the user privacy.

In this work, we aim to further explore the theoretical limits of reconstruction, speedup and stabilize the reconstruction procedure. We show that a single input may be reconstructed with the analytical form, regardless of network depth using a fully-connected neural network with one hidden node. Then we generalize this result to a gradient averaged over batches of size $B$. In this case, the full batch can be reconstructed if the number of hidden units exceeds $B$. For a \ac{cnn}, the number of required kernels in convolutional layers is decided by multiple factors, e.g., padding, kernel and stride size, etc. We require the number of kernels $h\geq (\frac{d}{d^{\prime}})^2C$, where we define $d$ as input width, $d^{\prime}$ as output width after convolutional layer, and $C$ as channel number of input. We validate our observation and demonstrate the improvements using bio-medical (fMRI, \ac{wbc}) and benchmark data (MNIST, Kuzushiji-MNIST, CIFAR100, ImageNet and face images).
\end{abstract}

Federated Learning  \citep{konevcny2016federated} was proposed by Google for distributed network training  and has created tremendous interest. One important virtue of \ac{fl} is to keep data on the generating-device to avoid information leakage due to direct data transmission. However, the distributed setup opens to exposure for attackers, and sensitive information may be disclosed in other forms. Recent work \citep{zhu2019deep} demonstrated a severe attack - reconstruction of input data within an \ac{fl} environment. This idea was similar to model inversion \citep{fredrikson2015model}; however, it is easy and effective to apply this attack in a federated setup. By imitating an honest server, data may be reconstructed with the access to gradient updates and model parameters. It seriously breaches the promise of \ac{fl} - data is kept locally. Here we aim to explore the limits and efficiency of this attack. We offer theoretical analysis of the reconstruction based on both a fully-connected neural network and \ac{cnn}, from a perspective of solving a linear system. Our main contributions are summarized as the follows.
\begin{itemize}
\itemsep0em
    \item We show that reconstruction of input amounts to solving a set of linear equations, and based on which we give minimal structural conditions for reconstruction based on the fully-connected neural network (a.k.a.\ac{mlp}) and \ac{cnn}.
    \item We show that \ac{mlp} only needs \textbf{a single} node in \textbf{one} hidden layer to reconstruct a single input image, regardless of the model depth. Complementing \cite{zhu2019deep}, we derive a closed form for a lossless reconstruction in this case. Moreover, we generalize the result to batch reconstruction and show that for a full reconstruction the number of hidden units has to exceed the batch size.
    \item We show that for \ac{cnn}, the number of kernels in every convolutional layer should be such that the size of output after passing through the convolutional layers exceeds the size of original input. 
    \item We propose a reconstruction method in response to three cases: 1) single-instance reconstruction using \ac{mlp}; 2) single-instance reconstruction using \ac{cnn}; 3) batch reconstruction.
    \item We also suggest to include a regularizer for batch reconstruction, which increases the numerical stability during iterative optimization and outputs more faithful reconstructions.
\end{itemize}

The paper is organized as follows: in Section \ref{sec:related}, we briefly introduce \ac{fl} and present key related works and in Section \ref{sec:method} we focus on our reconstruction method and offer the theoretical analysis based on the model architecture. Finally, we provide numerical demonstrations in Section \ref{sec:result} and conclude in Section \ref{sec:conclusion}.

\section{Background and related works}
\label{sec:related}
\subsection{Federated Learning}
\vskip -2mm
\ac{fl} is a distributed \ac{ml} paradigm that incorporates a set of distributed workers (customers) and a server to jointly train a global model. Unlike traditional Cloud-based ML \citep{yang2019federated}, it has no (training) data transmission between workers and the server. Instead, the set of workers collaborate to optimize the global cost function that is estimated by a collection of local cost functions. The workers own their data, the full copy of the global model, and implement local training (minimizing local cost functions) using their data and update the gradients with the server. Without loss of generality, we assume that each worker $i$ owns data generated from the empirical distribution $\hat{\mathscr{D}}_i$, and the data owned by different workers might be heterogeneous. Given $p$ active workers participate in each round, and $m_i$ is the amount of data on worker $i$ and $m$ is the sum of all data $m=\sum_{i=1}^p m_i$. All the workers share the same batch size $B$. The global empirical loss function $\hat{\ell}$ can be approximated by the weighted combination of local loss functions $\hat{\ell}_i$. It is defined as: 

\begin{equation} \label{global_eq}
\hat{\ell}(x,y) = \sum_{i=1}^p \frac{m_i}{m} \E_{(x,y)\sim \hat{\mathscr{D}}_i}[\hat{\ell}_i(f_w(x),y)]
\end{equation}

Each iteration consists of two stages: local training (workers) and aggregation (server). More specifically, after distributed workers have completed local training, they share the corresponding gradients with the server, and the server aggregates the gradients based on the given criterion and finally it sends the updated global model (or aggregated gradients) back to the workers. The workers will use the updated model for the next iteration. This procedure can be repeated for multiple times.

\subsection{Related work}
\vskip -2mm

The feasibility of input reconstruction based on the gradients and parameters of the neural works was demonstrated by \cite{zhu2019deep}. They proposed to reconstruct input from an initial guess sampled from normal distribution and iteratively optimize the reconstruction by minimizing the distance between gradients and guessed gradients. However, it does not give strong experimental results on larger batches. \cite{zhao2020idlg} expand on \cite{zhu2019deep} to improve the label prediction accuracy. \cite{geiping2020inverting} proposed a new cost function and reconstruct the input based on deep neural networks like ResNet \citep{he2016deep}. \cite{wei2020framework} introduced a framework for evaluation of privacy leakage and relation to \ac{fl} hyperparameters. \cite{pan2020theory} also focus on deep neural network like VGG \citep{simonyan2014very}, GoogLeNet \citep{szegedy2015going}, which are normally are more informative for the reconstruction. We are interested in minimal requirements analysis of reconstruction based on the network structure which is currently under-explored. 

\section{Reconstruction method and theoretical analysis}\label{sec:method}
\vspace{-2mm}
In this section, we will first present the reconstruction method and then introduce the theoretical analysis of minimal reconstruction conditions based on the two architectures: \ac{mlp} and \ac{cnn}.

\subsection{Reconstruction method}
\vskip -2mm
Input reconstruction is essentially an inverse problem, without the loss of generality, say we have any function $G$ parameterized by $w$, which maps from input $x\in R^d$ to a (gradient) vector $v$, thus $G_w:\rbb^{d}\to v$. For the batch case, the expected gradient is $\overline{v} = \frac{1}{B}\sum_{i=1}^B G(x_i,y_i;w) $. We aim to compute $x$ with the knowledge of model parameters $w$ and output $v$ (or $\overline{v}$). If $G^{-1}$ exists, we can compute $x$ directly, which is typically not the case. The intriguing question is that is it still possible to reconstruct $x$ in the noninvertible case? We show that in Section \ref{full_nn}, one-instance reconstruction based on \ac{mlp} has an analytical form and can be computed directly. This is different from \cite{zhu2019deep} where they consider it as noninvertible and address it as an optimization problem. For one-instance \ac{cnn} reconstruction, we propose a two-step method where we first compute the output of convolutional layer using the closed-form mentioned before, and then we apply the optimization method. While, for batch reconstruction, there is no analytical form in most cases. We convert it to the optimization problem; more specifically, we start from a random guess $(\hat{x},\hat{y})$, and gradually pull it to close to the original input $x$ by minimizing the cost function, like proposed by \cite{zhu2019deep}. For the batch reconstruction, we augment the cost function by an additive regularizer as $L(.)+\lambda R(.)$ where $L(.)$ is the distance between ground-truth gradients and guessed gradients. For $R(.)$ we suggest an orthogonality regularizer defined as $R = \lambda \sum_{k\neq k^{\prime}=1}^n \left(\hat{x}_k^{\intercal} \hat{x}_{k^{\prime}}\right)^2$, if $\hat{x}_k$ and $\hat{x}_{k^{\prime}}$ are orthogonal the product is zero, which implies that the optimizer promotes solutions where the batch members are dissimilar. Or a L2 regularizer defined as $\lambda \sum_i \hat{x}_i^{\intercal}\hat{x}$. We experimentally explore the critical role of the regularizer for large batch size in Section~\ref{sec:result}.

\begin{algorithm}[ht]
\caption{Insecure \ac{fl} with reconstruction attack}
\label{Alg_FL}
\begin{algorithmic}[1]
\State \text{\textbf{Initialization:} $w^0$}
\For{t=1,...T}
\State \textbf{$=>$workers:}
\For{j=1,2,..,p} \text{$p$ workers (in Parallel)}
\State \text{ $v^t_j = \nabla\hat{\ell}_j (f(X_{j}^t;w^t),Y_{j}^t)$}
\State \text{with $(X_{j}^t=\{x_{jk}^t\}_{k=1,..B},Y_{j}^t=\{y_{jk}^t\}_{k=1,...,B}) \sim \hat{\mathscr{D}}_j$}
\State \text{ share $v_{j}^t$ with server}
\EndFor
\State \textbf{end}
\State \textbf{$=>$server(attacker):}
\State \text{$w^{t+1} = w^t - \eta\times \frac{m_j}{m} \sum_{j=1}^p v_j^t$}
\State \text{share $w^{t+1}$ with workers for next round}
\State $\hat{X}^t_j,\hat{Y}^t_j=$\textit{Reconstruction}$(v_j^t,w^t,m,\lambda,\text{prior})$(Algorithm \ref{our_rec})
\EndFor
\State \textbf{end}
\end{algorithmic}
\end{algorithm}

We show insecure \ac{fl} with potential reconstruction attack in Algorithm \ref{Alg_FL} where the attacker could be the server who has the complete knowledge of gradients update and model parameters from each worker $j$ at each round $t$. In Algorithm \ref{our_rec}, we present the reconstruction approaches in response to three cases: 1) one-instance \ac{mlp} reconstruction, 2) one-instance \ac{cnn} reconstruction, and 3) batch reconstruction (using \ac{mlp} and \ac{cnn}). The pseudo code in Algorithm \ref{itr_opt} demonstrates the iterative optimization step. 

\begin{algorithm}[ht]
\caption{\textit{Reconstruction}}
\label{our_rec}
\begin{algorithmic}[1]
\State \text{Input: $v,w,m,\lambda,\text{prior}$}
\If{1) single recon. $\&$ mlp}\Comment{case one}
\State \text{\textbf{Return} $\hat{x}=\frac{\partial \ell}{\partial w^{1}_{1i}}/\frac{\partial \ell}{\partial b_1^1}  \quad \forall i<d$} \Comment{$w^1_{1i}, b^1_1$ are the weights and bias in $1^\text{st}$ hidden layer, $d$ is input dimension}
\EndIf
\State \textbf{End}
\If{2) single recon. $\&$ cnn}\Comment{case two}
\State \text{$\hat{z}=\frac{\partial \ell}{\partial w^{1}_{1i}}/\frac{\partial \ell}{\partial b_1^1} \quad \forall j < d^{\prime}$}\Comment{$w^1_{1j}, b^1_1$ are the weights and bias on $1^{\text{st}}$ hidden layer after conv. layer, $d^{\prime}$ is dimension of output of conv. layer}
\State \textbf{Return} $\text{Itr}\_\text{rec}(\hat{z},w_{\text{partial}},"\text{partial}",m,\lambda,\text{prior})$\Comment{$w_{\text{partial}}$ refers to the params. of convol. layer}
\EndIf
\State \textbf{End}
\If{3) batch reconstruction}\Comment{case three}
\State \textbf{Return} $\text{Itr}\_\text{rec}(v,w,"\text{all}",m,\lambda,\text{prior})$
\EndIf
\State \textbf{End}
\end{algorithmic}
\end{algorithm}

\begin{wrapfigure}{L}{0.5\textwidth}
\vspace{-6mm}
\begin{minipage}{0.5\textwidth}
 \begin{algorithm}[H]
    \caption{\textit{$\text{Itr}\_\text{rec}$}}\label{itr_opt}
  \begin{algorithmic}[1]
\State \text{Input: $v,w,\text{flag},m,\lambda,\text{prior}$}
\For{i=0,1,2,..,I} \text{I iterations}
\If {i==0}
\If {prior==uniform}
\State \text{$\hat{X}_0,\hat{Y}_0 \sim \ubb(\mathbb{0},\mathbb{1})$}
\EndIf
\If {prior==normal}
\State \text{$\hat{X}_0,\hat{Y}_0 \sim \nbb(\mathbb{0},\mathbb{1})$}
\EndIf
\State \text{$\hat{v}_1= G(\hat{X}_0,\hat{Y}_0;w)$}\Comment{$G$ is function of computing gradients}
\Else
\State \text{$\hat{v}_i = G(\hat{X}_i,\hat{Y}_i;w)$}
\If{flag=="all"}
\State \text{$L= \norm{v-\hat{v}_i}^2_2+\lambda R(\hat{X}_i)$}\Comment{$R$ is the regularizer}
\Else
\State \text{$L= \norm{v-\hat{v}_i}^2_2$}
\EndIf
\State \textbf{end}
\State \text{update: $\hat{X}_{i+1}=\hat{X}_{i}-\eta \times \nabla_{\hat{X}_{i}} L$}
\State \text{\hspace{1.2cm}$\hat{Y}_{i+1}=\hat{Y}_{i}-\eta \times \nabla_{\hat{Y}_{i}} L$}
\EndIf
\If{(i//m)==0}
\State \text{$\lambda=0.9*\lambda$}
\EndIf
\State \textbf{end}
\EndFor
\State \textbf{end}
\State return $\hat{X}_I,\hat{Y}_I$
\end{algorithmic}   
\end{algorithm}
\end{minipage}
\vspace{-12mm}
\end{wrapfigure}
Our method implementation deviates from the pioneering work \citep{zhu2019deep} in a few ways. First, we derive a closed-form for one-instance \ac{mlp} reconstruction, which leads to a faster and more accurate reconstruction (almost lossless). Second, we divide one-instance \ac{cnn} reconstruction into two steps; first, we directly compute the output of convolutional layer using the advantage of closed-form and based on which we reconstruct the input (a.k.a deconvolution), which speeds up the overall reconstruction. Last, we expand the cost function with either an orthogonality regularizer or L2 regularizer for batch reconstruction. The orthogonality regularizer may penalize the similarities between reconstructed images (since we only know the average gradient of the batch), mainly when image patterns are similar (e.g., MNIST).  In contrast, L2 regularizer offers faster convergence, in particular, when batch size is large. Our main focus in this work is to explore and validate the minimal condition of the full reconstruction.

\subsection{Reconstruction with fully-connected neural network}\label{full_nn}
\vskip -2mm
Say we have a one-layer \ac{mlp} $f$, with $n_1$ units in hidden layer and $n_2$ units in output layer (if classification task, $n_2$ is equal to the number of classes). Thus, we can define the output of model as  $a_j = f_w(x)=\sum_{i=1}^{n_1}w^2_{ji}\sigma(w^1_ix+b^1_i)+b^2_j \quad \forall j\in [1,n_2]$, where $w^1$ and $b^1$ are the weights and bias in hidden layer, and $w^2$, $b^2$ in output layer and $\sigma(x)$ is sigmoid (monotonic) activation function. For the classification task, we employ cross-entropy as the cost function $\ell(p_i,y_i)=-\sum_j^C y_{ij}\log p_{ij}$ where $p_{ij}$ is the output of softmax function, and $y_i$ is the one-hot encoding vector with all zeros except the corresponding class indicating one. 

\vskip -2cm

\begin{proposition}[\textsc{one-instance \ac{mlp} reconstruction}]
\label{one-mlp}
To reconstruct one input based on \ac{mlp}, we derive the analytical form to compute the (almost) lossless input, with only \textbf{single} unit in the first hidden layer as long as bias term exists, regardless how deep the network is.
\end{proposition}

The comprehensive proof can be found in appendix~\ref{proof-mlp1}. It can be generalized to deep fully-connected neural network (say last layer is $L$). We only need two ingredients in our recipe, the partial derivative w.r.t the bias term and weights in the first hidden layer (after input layer). It shows that training with \ac{sgd} is very vulnerable to reconstruction attack in FL (distributed ML), in particular, on fully-connected neural network no matter how deep it is.

\begin{proposition}[\textsc{batch \ac{mlp} reconstruction}]
\label{batch-mlp}
For batch reconstruction, the number of units in first hidden layer should meet $n_1 \geq B$, given the high-dimension input whose dimension $d\gg n_2, d\gg B$.
\end{proposition}
\vspace{-5mm}
\begin{proof}
(sketch:) Reconstructing $x_1, x_2, ..., x_B \in \rbb^d$ approximates to solving the linear equations. Given an invertible sigmoid function and a single output $y_i$, we may compute the unique $\sigma^{\prime}(x_i)$. The number of equations exceeds the number of variables $n_2+n_1n_2+n_1+n_1d \geq Bn_2+n_1B+Bd$, thus $n_1\geq \frac{Bn_2+Bd-n_2}{n_2+1+d-B}$. Typically we have $d\gg n_2, d\gg B$, and it is dominated by $\frac{B(d+n_2)}{d+n_2}$, therefore $n_1\geq B$.
\end{proof}
 \vskip -4mm
In general, batch reconstruction is more challenging since we typically only know the average or the sum of the gradients and aim to reconstruct every individual instance. We refer to this procedure as \textit{demixing} in the following context. Theoretically, if the number of equations is identical or greater than batch size and all the equations are independent, it is solvable. However, it is not easy to solve in practice, mainly when $x$ is high-dimensional and the scales in the linear system are small. We use an iterative method to solve it. Besides, the to-be-optimized input value during optimization might introduce the saturation of sigmoid function, i.e., $\sigma(x)\approx 0, \sigma(x)\approx 1$ regardless the change of $x$ outside the interval $[-4,4]$. The choice of optimization method is important. For a high-dimensional (non-sparse) input, second-order or quasi-second-order methods sometimes fail since they are designed to search for the zero-gradient area. The number of saddle points exponentially increases with the number of input dimensions \citep{dauphin2014identifying}. The first-order method, e.g., Adam \citep{kingma2014adam} takes much more iterations to converge, but it is relatively more stable, likely to escape from the saddle points \citep{goodfellow2016deep}.

\subsection{Reconstruction with convolutional neural network}
\vskip -6mm
\begin{proposition}[\textsc{single-layer \ac{cnn} reconstruction}]
\label{cnn-theorem}
To reconstruct a single-layer \ac{cnn} immediately stacked by a fully-connected layer, $h\geq (\frac{d}{d^{\prime}})^2C$ kernels are required, where $C$ is the channel number of input, $d$ is the width of input, and $d^{\prime}$ is width after convolutional layers.
\end{proposition}\label{prop_cnn}
\vskip -2mm
For a single convolutional layer \ac{cnn}, the kernel parameters (kernel size $k$, padding size $p$, stride size $s$) determine the output size $d^{\prime}$ after convolutional layers. Say we have input $X\in \rbb^{B\times C\times d\times d} $, for the simplicity we assume height and width are identical, and $C$ is the channel number and $B$ is the batch size. We define a square kernel with width $k$ (weights indicated as $l^0$), bias term $r$, and we have $h$ kernels. After convolutional layer, the width of output is $d^{\prime}=\frac{d+2p-k}{s}+1$. 
The output of convolutional layer is defined in eq. (\ref{conv_operator}). 
\begin{equation}\label{conv_operator}
\begin{split}
        z_{\underline{m}ij}&=(\sum_{c=1}^C\sum_{g=1}^{k}\sum_{n=1}^{k}l_{\underline{m}c gn}^{0}\hat{x}_{c,si+g-1,sj+n-1})+r_{\underline{m}}\\
    & \forall(i,j)\in [1,d^{\prime}]\times[1,d^{\prime}], \forall \underline{m}\in[1,h]
\end{split}
\end{equation}
We include the proof in appendix~\ref{proof-cnn}.
The number of equations should be equal or greater than the number of unknowns. From eq. (\ref{conv_operator}), we have $(d^{\prime})^2Bh$ equations and $d^2BC$ unknowns, thus we need $h\geq (\frac{d}{d^{\prime}})^2C$, with the assumption that $H$ is known, which can be solved by enough units in dense layer from Proposition \ref{batch-mlp}. One special case is that convolutional layer is stacked by an output layer directly (no dense layer), then we need to meet $h\geq (\frac{d}{d^{\prime}})^2CB$, $n_0 \geq \frac{n_1(B-1)}{n_1-B}$, and $1<B<n_1$ to solve $H$ since $\sigma$ is monotonic activation function ($n_0=h(d^{\prime})^2$). Note here $n_1$ indicates the number of units in output layer (as no dense layer). It can also be generalized to multiple convolutional layers, more details can be found in appendix~\ref{proof-cnn}.

\ac{cnn} reconstruction can be seen as a two-stage reconstruction: \textit{demixing} and \textit{deconvolution}. Namely, the demixing stage is essentially an inverse procedure of fully-connected network and we aim to reconstruct the output of convolutional layer (input of fully-connected layer). The deconvolution stage we want to reverse the convolutional step, starting from the output of convolutional layer to reconstruct the original input. For the demixing stage, we can apply the conclusion from \ac{mlp}, as long as the number of hidden units is equal or greater than the batch size, then theoretically we may evaluate the individual instance.

\vskip -2mm
\section{Experimental results}\label{sec:result}
\vskip -3mm
\textbf{Dataset and Setup.} MNIST and KMNIST are the handwritten digits and Kuzushiji accordingly, with size $1\times28\times28$ and it totally contains 10 classes. CIFAR100 contains 100 classes and every class has 600 images with size $3\times32\times32$. ImageNet contains 1000 classes, and each image has size $3\times64\times64$. Every image in \ac{wbc} dataset has size $3\times 240\times 320$ and four classes (Eosinophi, Neutrophil, Lymphocyte and Monocyte). We have implemented our method in Python and run it on a Titan X GPU with Architecture Maxwell and 12 GB of Ram.
\begin{figure}
    \centering
    \subfloat[ori.\label{subfig1:imagenet_ori}]{%
       \includegraphics[width=0.11\textwidth]{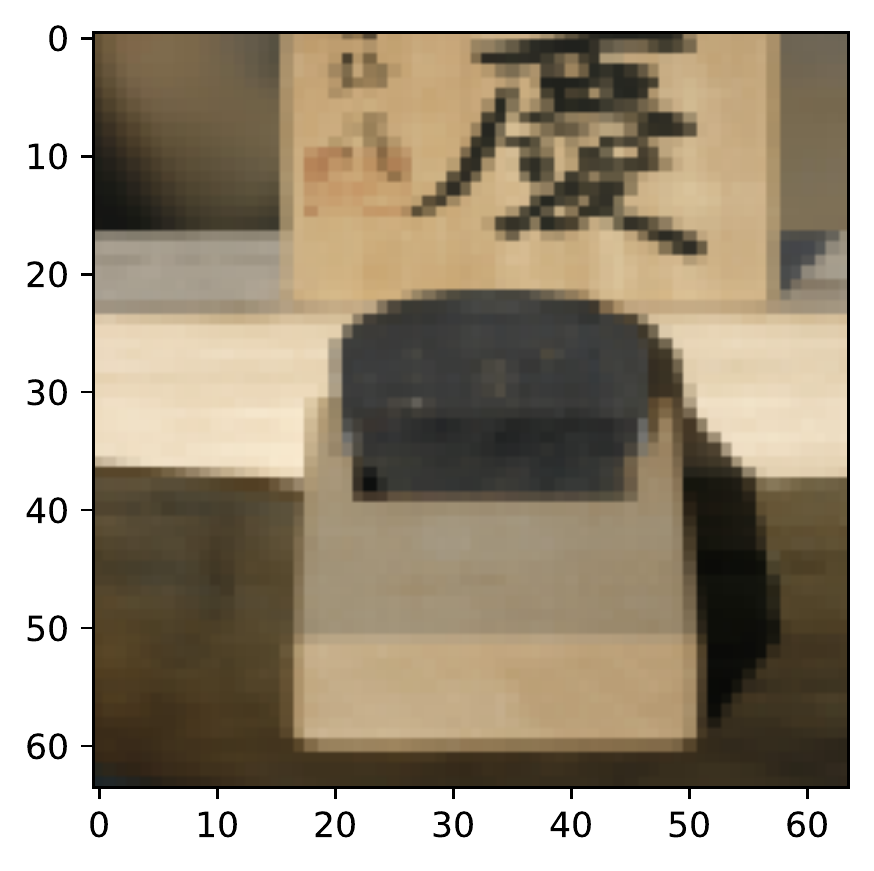}
     }
    \subfloat[ori.\label{subfig2:imagenet_ori}]{%
       \includegraphics[width=0.11\textwidth]{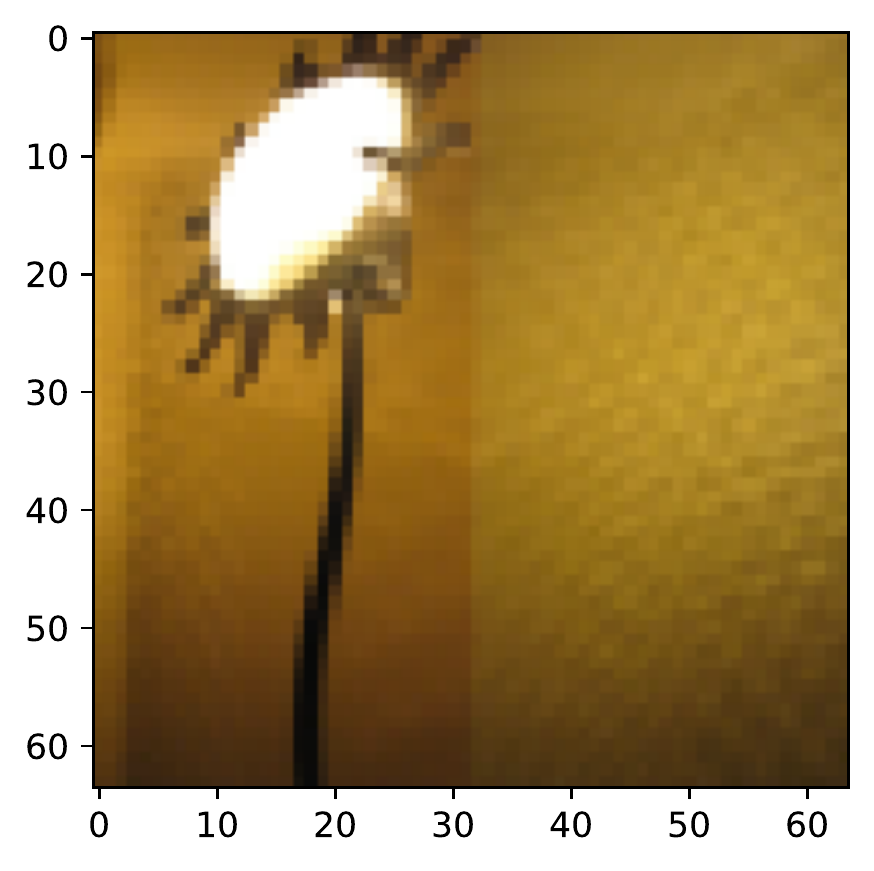}
     }
     \subfloat[ori.\label{subfig3:imagenet_ori}]{%
       \includegraphics[width=0.11\textwidth]{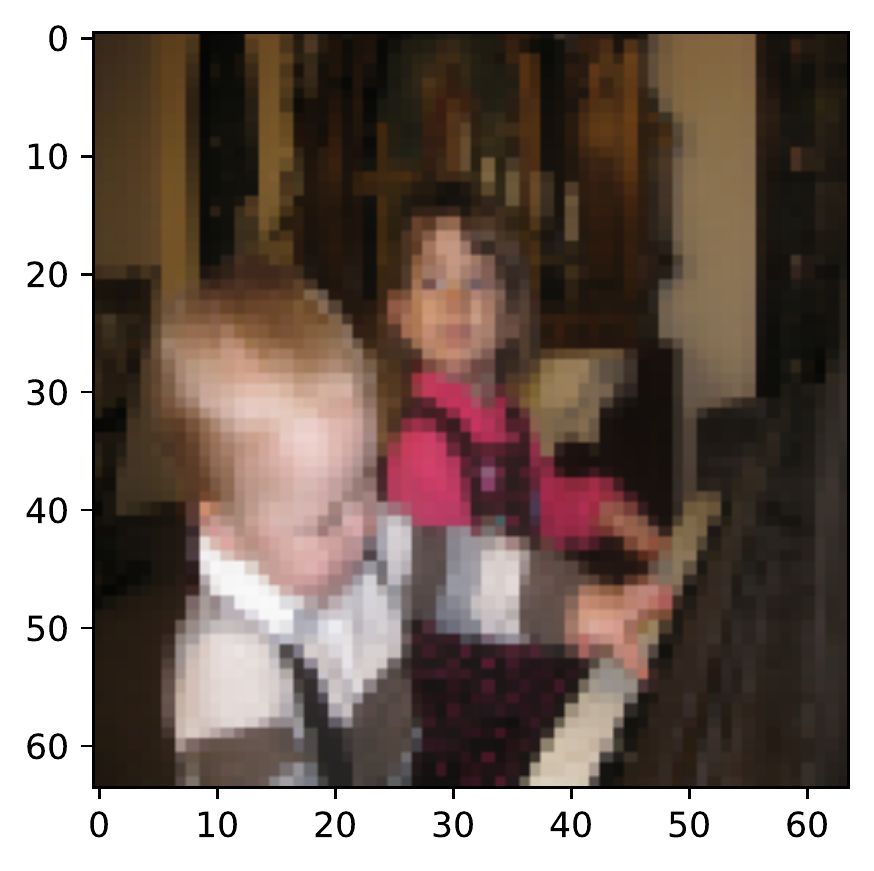}
     }
     \subfloat[ori.\label{subfig4:imagenet_ori}]{%
       \includegraphics[width=0.11\textwidth]{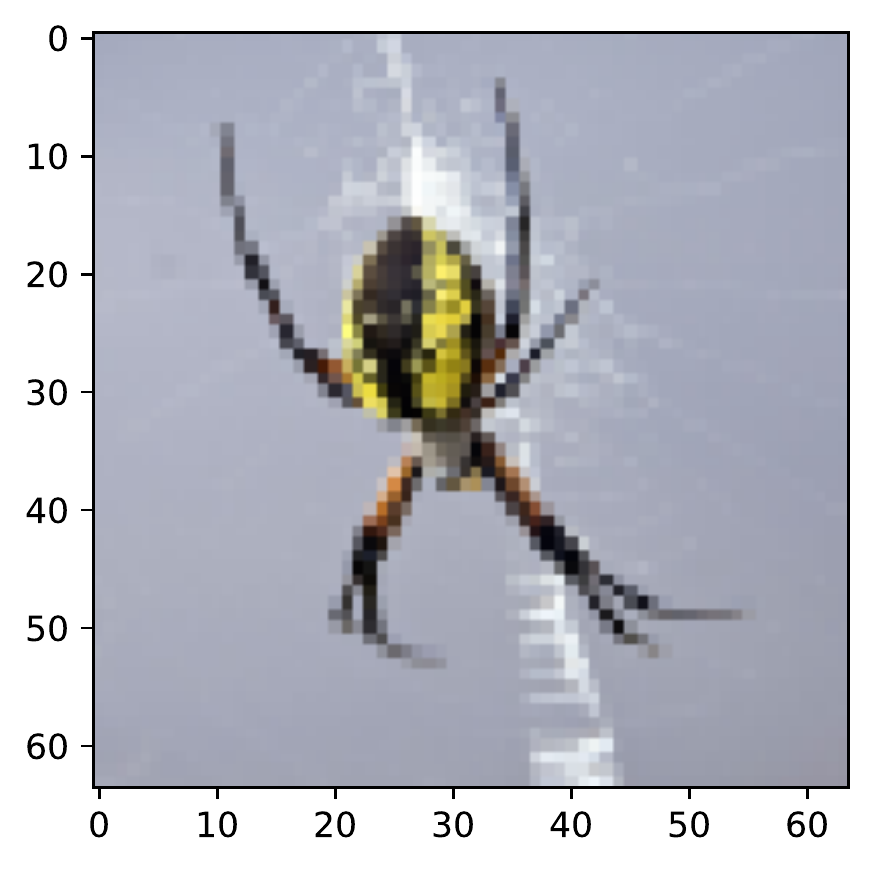}
     }
      \subfloat[ori.\label{subfig1:wbc_ori}]{%
      \includegraphics[width=0.11\textwidth]{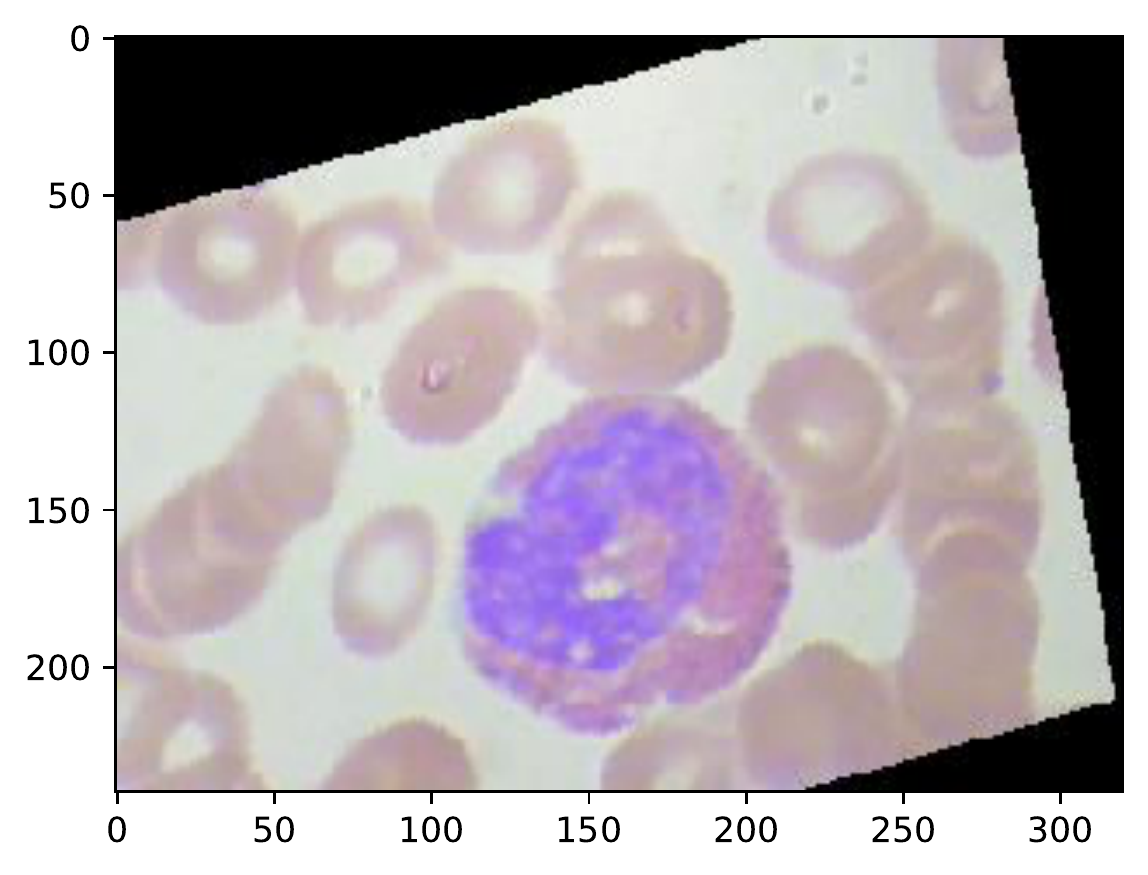}
     }
    \subfloat[ori.\label{subfig2:wbc_ori}]{%
      \includegraphics[width=0.11\textwidth]{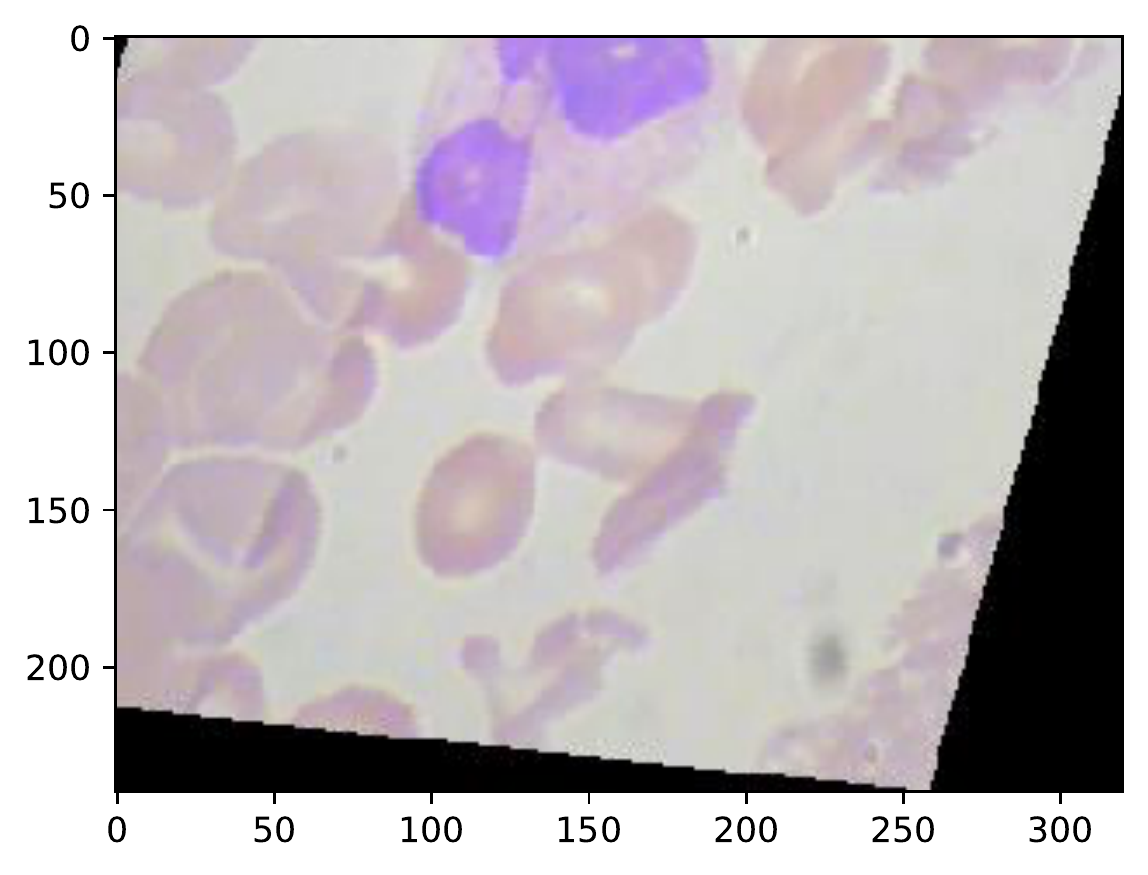}
     }
    \subfloat[ori.\label{subfig3:wbc_ori}]{%
      \includegraphics[width=0.11\textwidth]{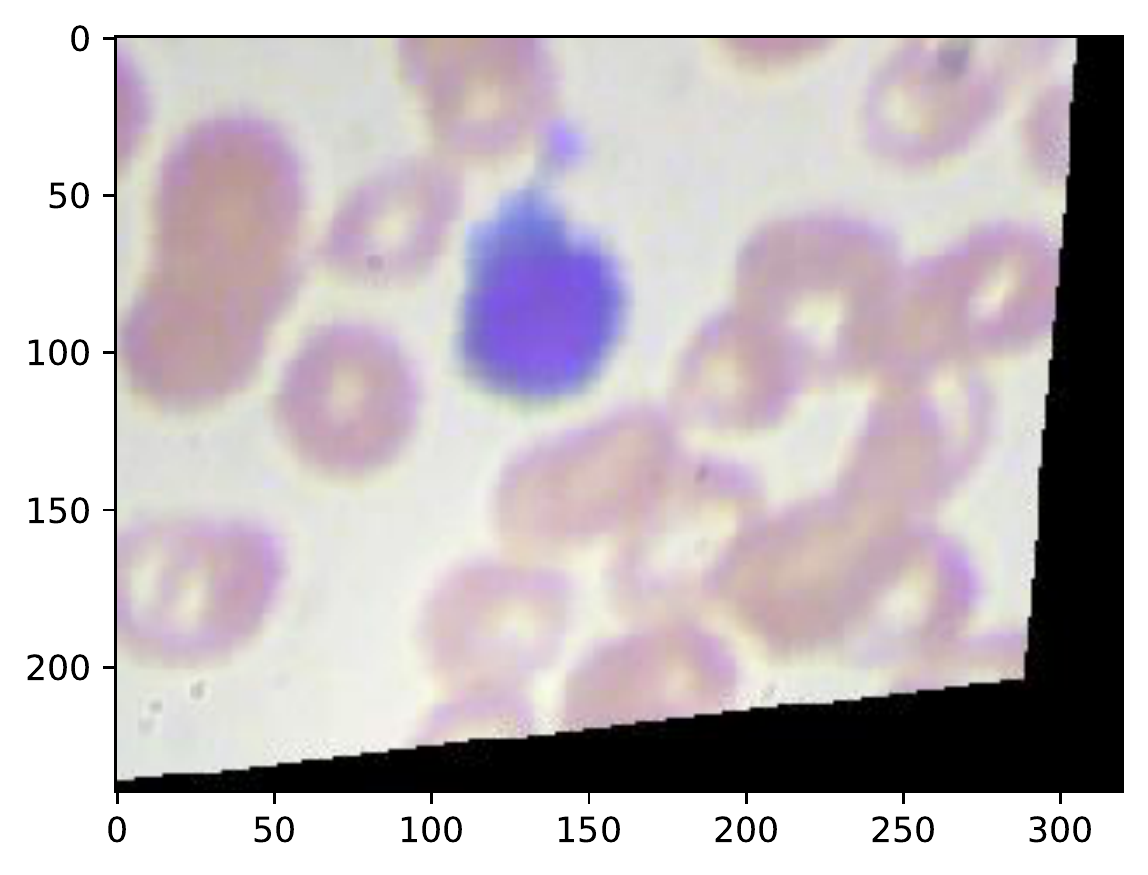}
     }
    \subfloat[ori.\label{subfig4:wbc_ori}]{%
      \includegraphics[width=0.11\textwidth]{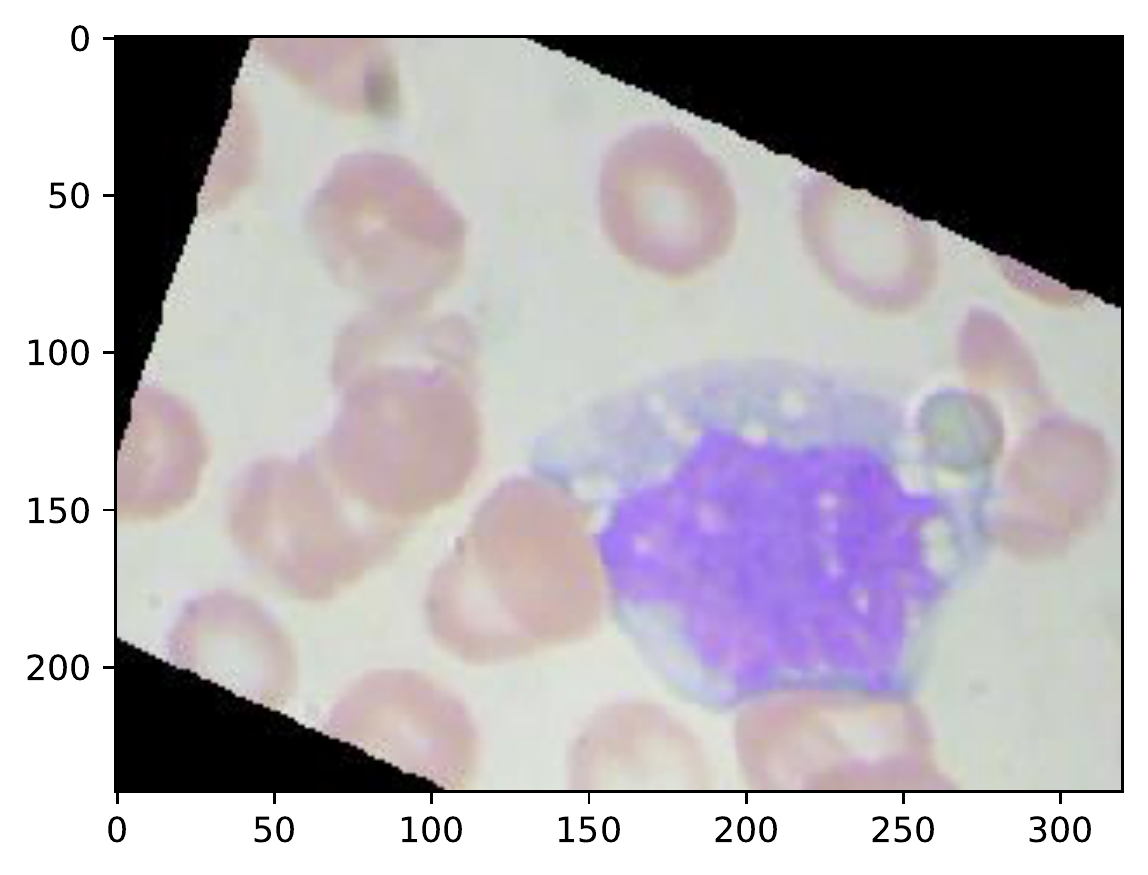}
     }
     
    \subfloat[rec.\label{subfig1:imagenet_rec}]{%
       \includegraphics[width=0.11\textwidth]{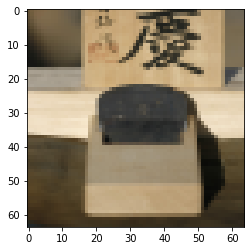}
     }
    \subfloat[rec.\label{subfig2:imagenet_rec}]{%
       \includegraphics[width=0.11\textwidth]{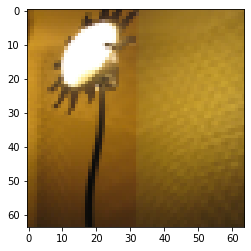}
     }
     \subfloat[rec.\label{subfig3:imagenet_rec}]{%
       \includegraphics[width=0.11\textwidth]{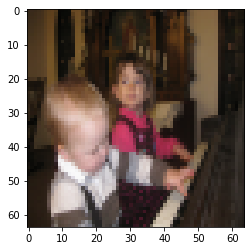}
     }
     \subfloat[rec.\label{subfig4:imagenet_rec}]{%
       \includegraphics[width=0.11\textwidth]{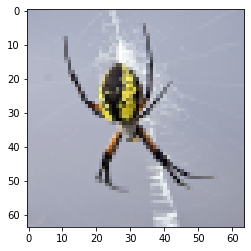}
     }
      \subfloat[rec.\label{subfig1:wbc}]{%
      \includegraphics[width=0.11\textwidth]{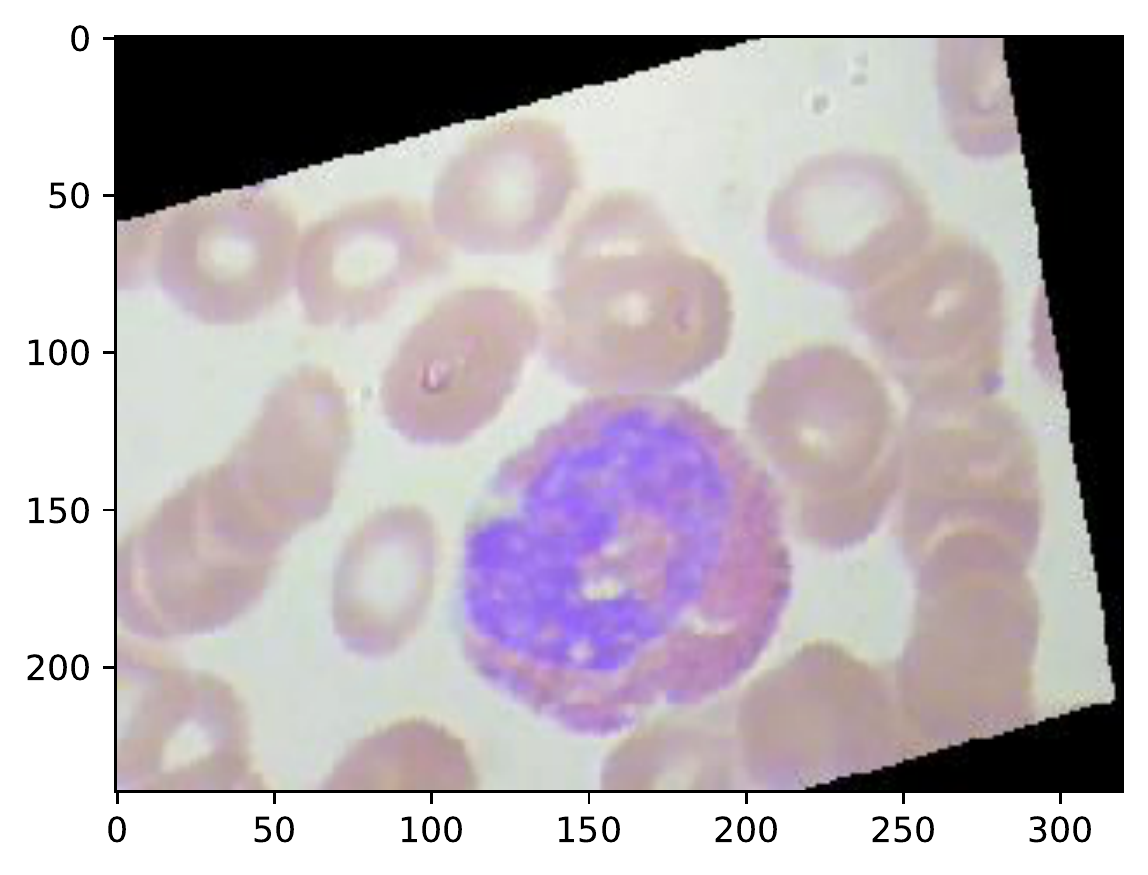}
     }
    \subfloat[rec.\label{subfig2:wbc}]{%
      \includegraphics[width=0.11\textwidth]{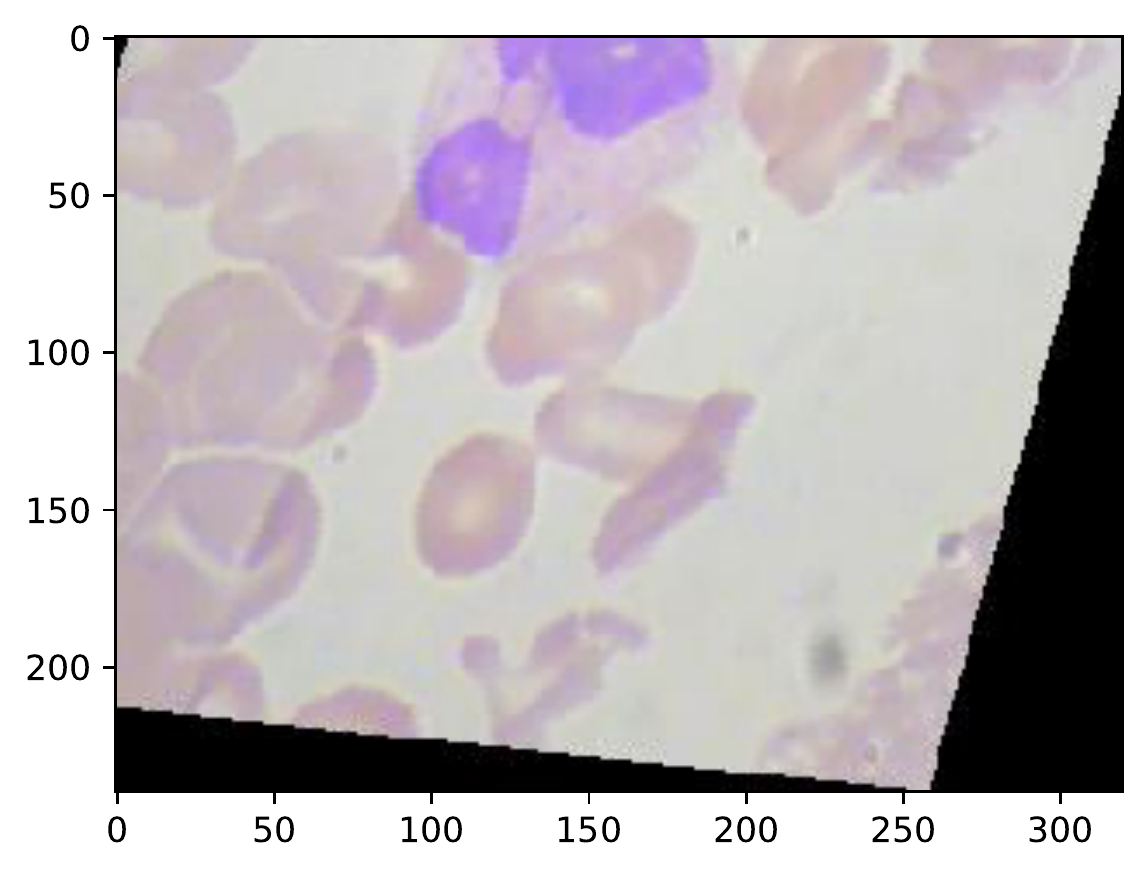}
     }
    \subfloat[rec.\label{subfig3:wbc}]{%
      \includegraphics[width=0.11\textwidth]{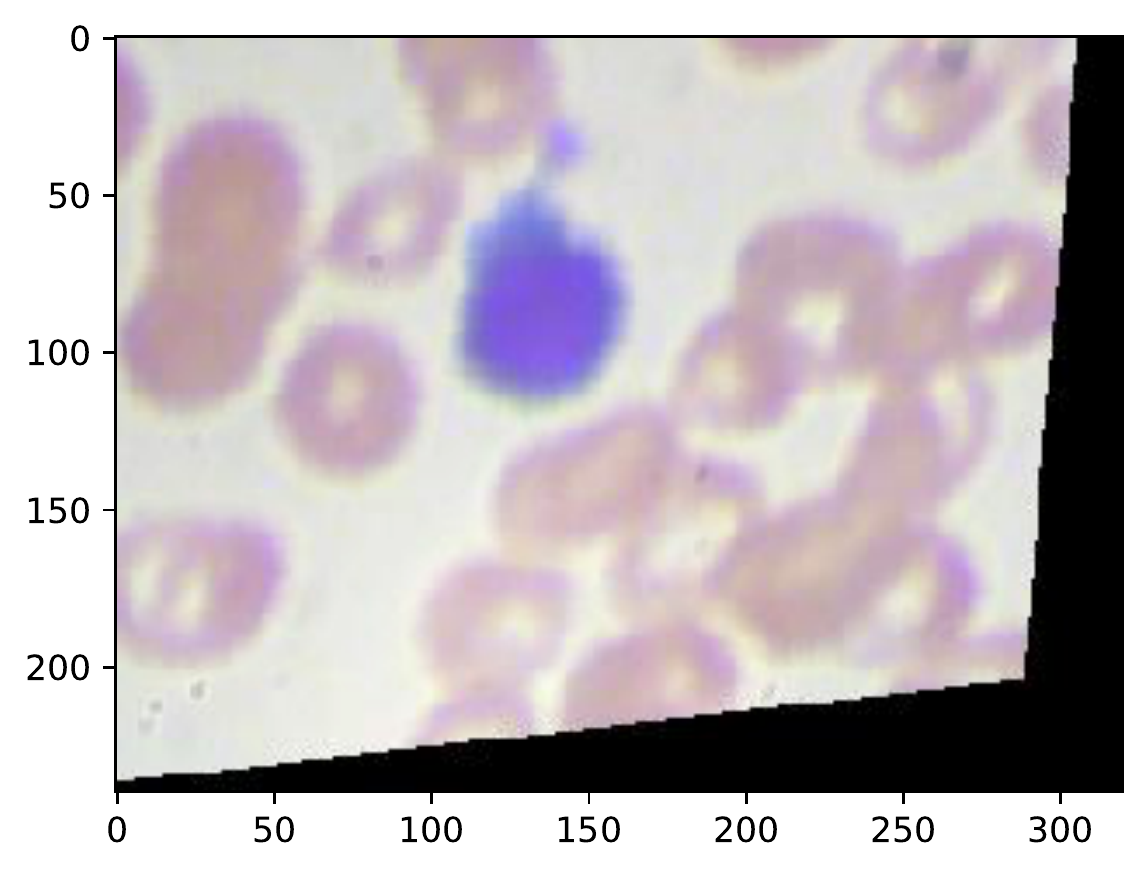}
     }
    \subfloat[rec.\label{subfig4:wbc}]{%
      \includegraphics[width=0.11\textwidth]{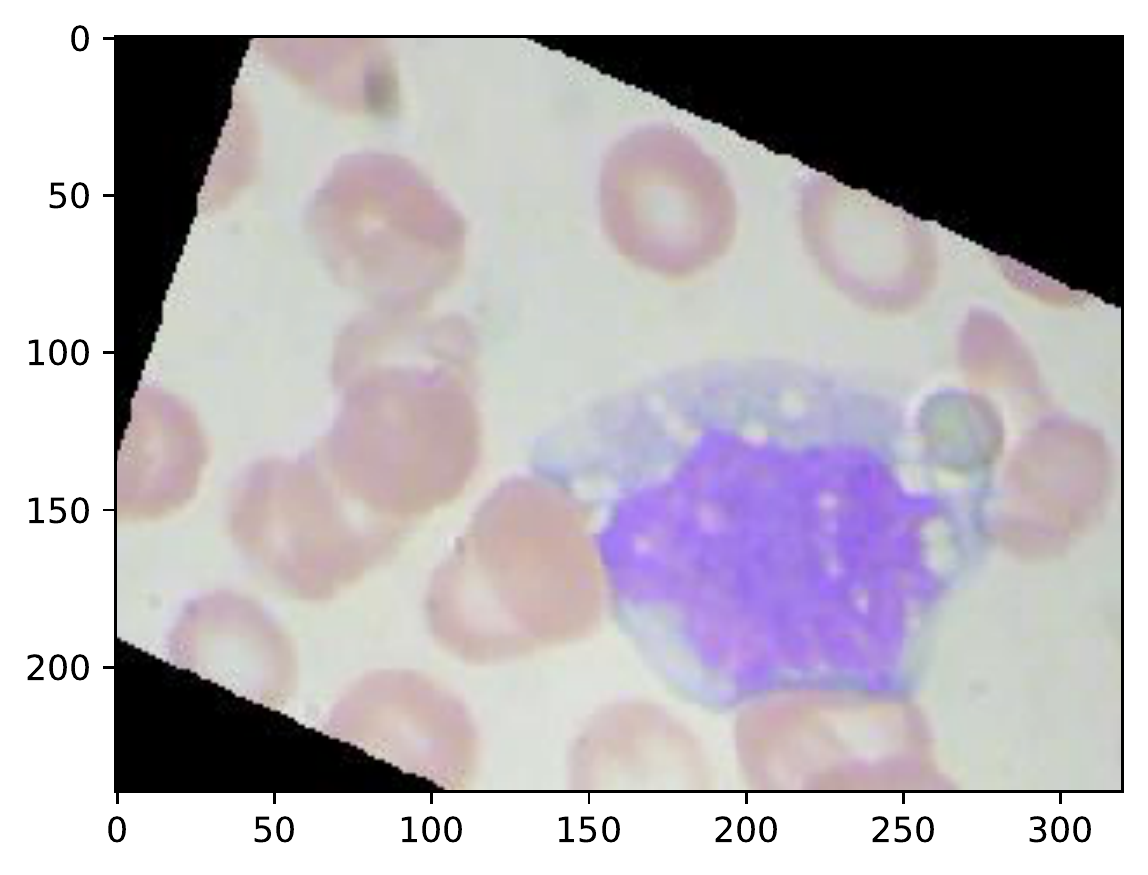}
     }
    \caption{ImageNet and White Blood Cell analytical reconstruction. The first row includes the original inputs and the second row corresponds to all the reconstructions.}
    \label{imagenet-single}
    \vskip -0.5cm
\end{figure}

\begin{figure}[!ht]
\centering
\subfloat[No regularizer(\citep{zhu2019deep})\label{mnist:noregularizer}]{%
       \includegraphics[width=0.4\textwidth]{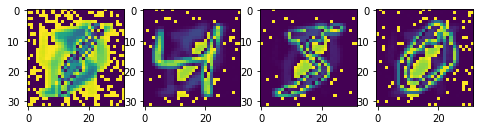}
     }
\subfloat[with orthogonality regularizer\label{mnist:regularizer}]{%
       \includegraphics[width=0.4\textwidth]{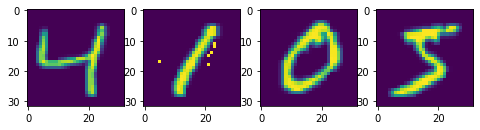}
     }
\vskip -1mm
\subfloat[1/4 recon. procedure\label{mnist:regularizer-1}]{%
       \includegraphics[width=0.24\textwidth]{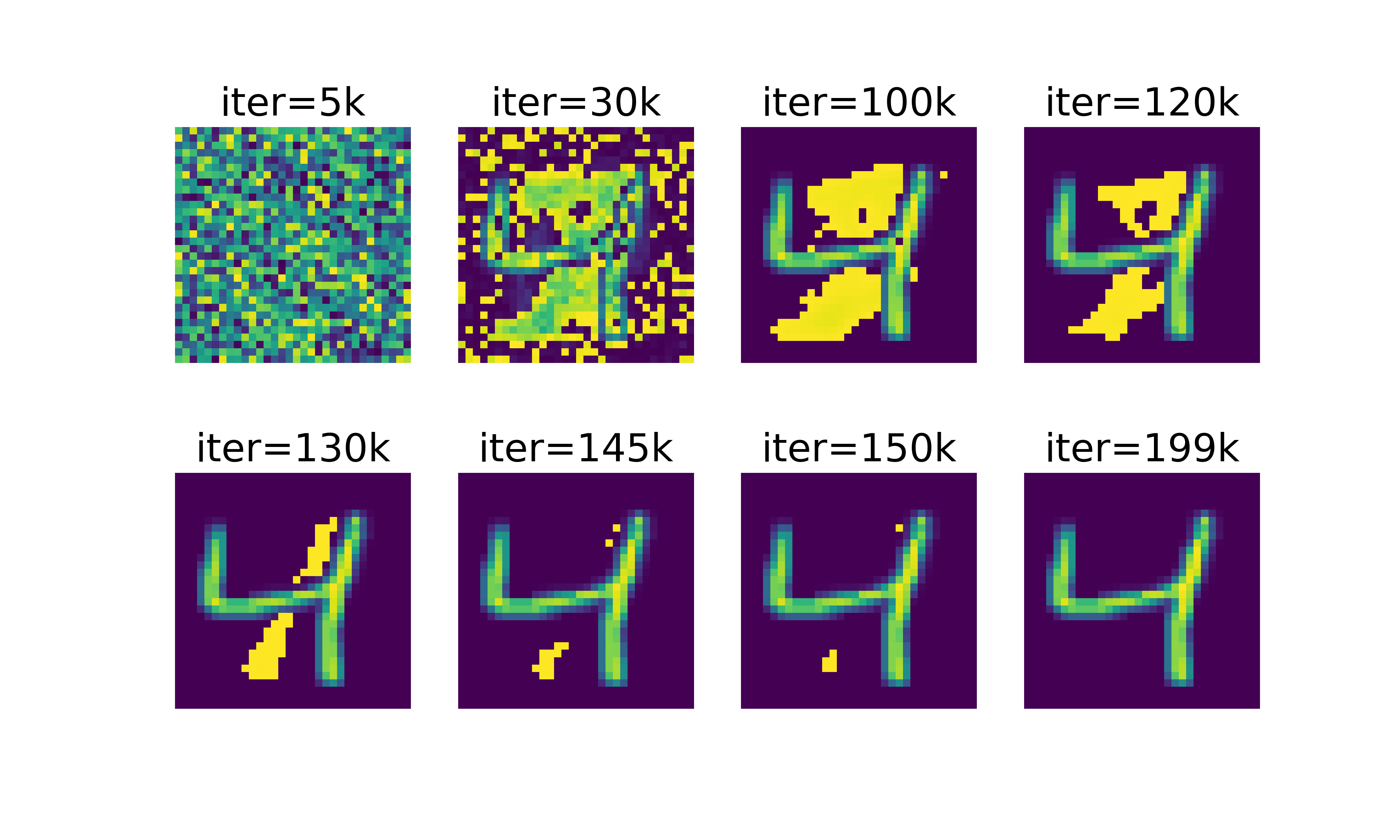}
     }
     \hfill
\subfloat[2/4 recon. procedure\label{mnist:regularizer-2}]{%
       \includegraphics[width=0.24\textwidth]{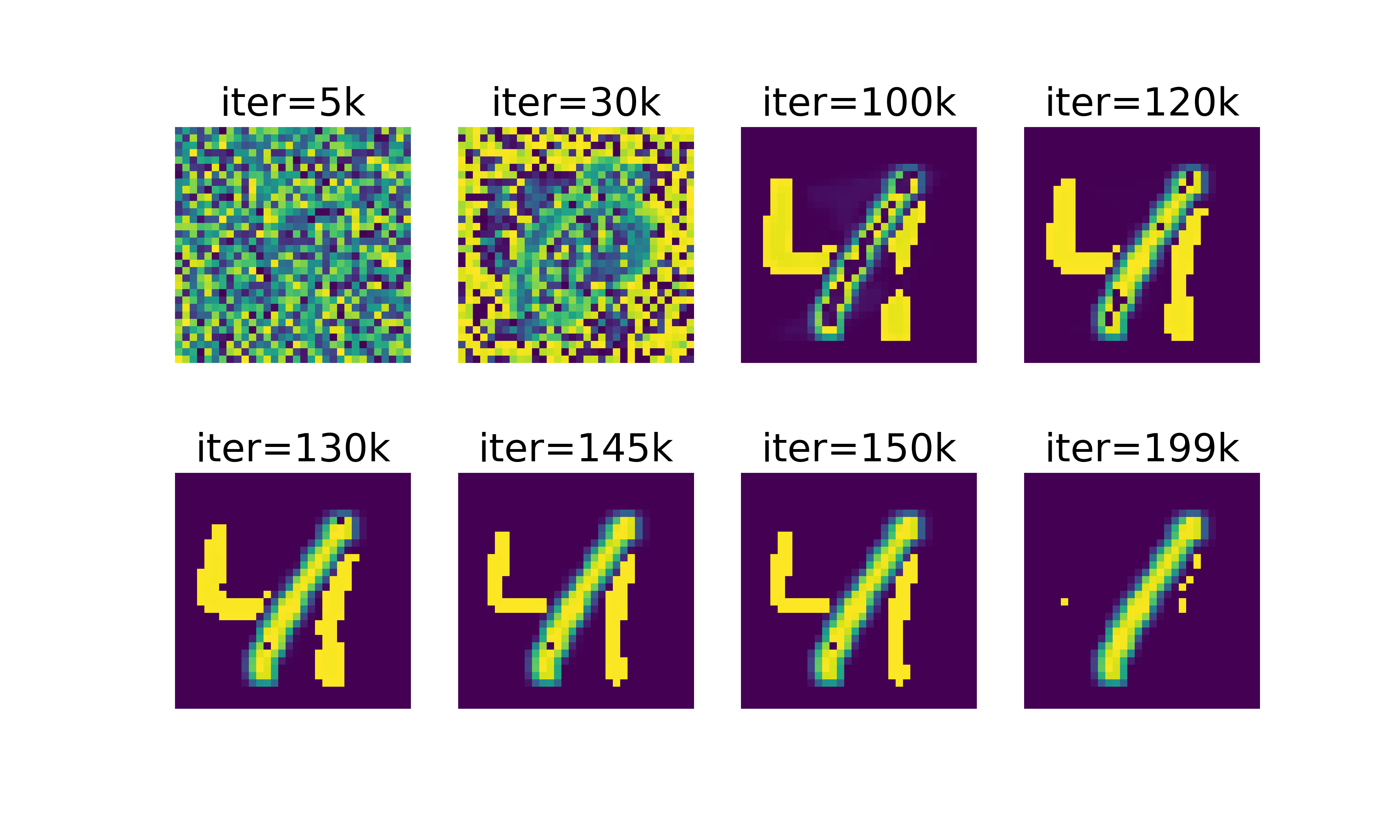}
     }
     \hfill
\subfloat[3/4 recon. procedure\label{mnist:regularizer-3}]{%
       \includegraphics[width=0.24\textwidth]{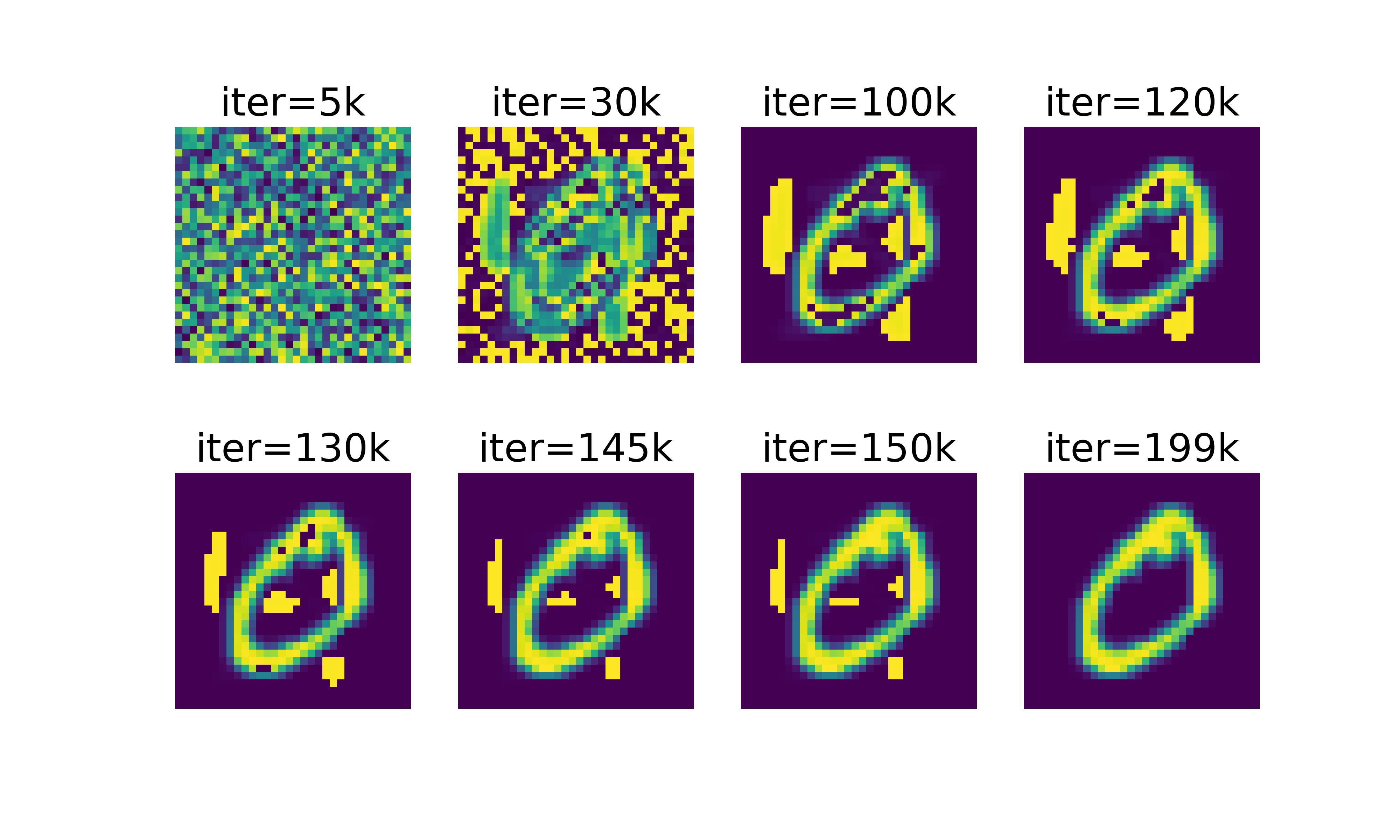}
     }
     \hfill
\subfloat[4/4 recon. procedure\label{mnist:regularizer-4}]{%
       \includegraphics[width=0.24\textwidth]{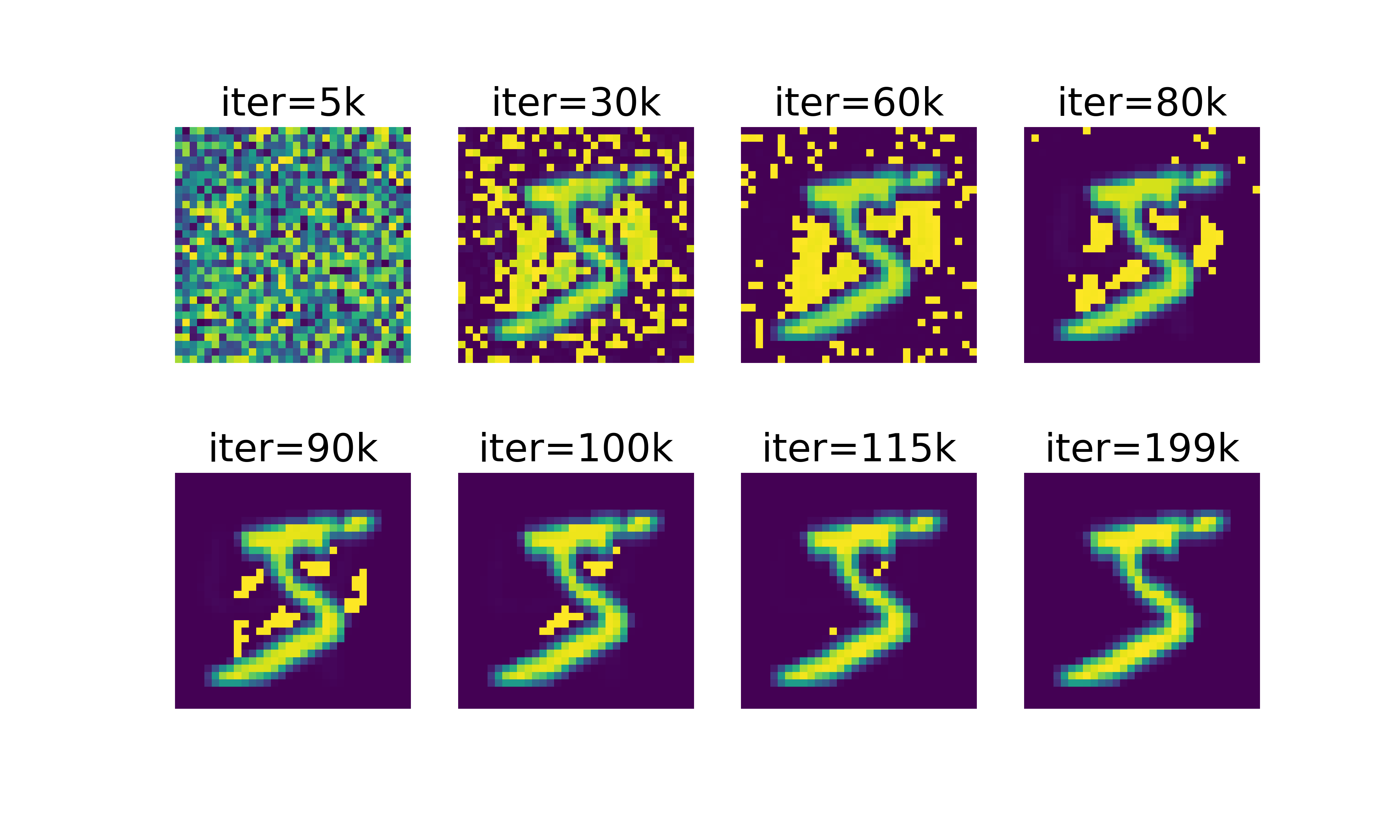}
     }
\caption{\textit{MNIST}: \ref{mnist:noregularizer} and \ref{mnist:regularizer} show the final reconstruction without and with regularizer accordingly. Moreover, the reconstructions procedures of \ref{mnist:regularizer} are separately shown in \ref{mnist:regularizer-1}, \ref{mnist:regularizer-2}, \ref{mnist:regularizer-3} and \ref{mnist:regularizer-4}.}
\label{fig:regularizer}
\end{figure}

\begin{figure}[!ht]
\vskip -3mm
    \centering
    \subfloat[1 filter\label{cifar100-onelayer1}]{%
    \includegraphics[width=0.09\textwidth]{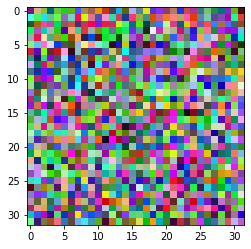}
     }
    \hfill
    \subfloat[5 filt.s\label{cifar100-onelayer2}]{%
    \includegraphics[width=0.09\textwidth]{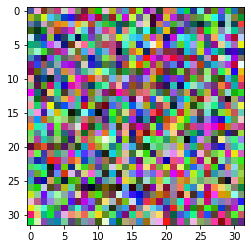}
     }
     \hfill
     \subfloat[11 filt.s\label{cifar100-onelayer3}]{%
    \includegraphics[width=0.09\textwidth]{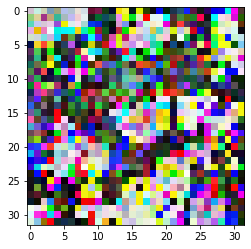}
     } 
     \hfill
     \subfloat[12 filt.s\label{cifar100-onelayer4}]{%
    \includegraphics[width=0.09\textwidth]{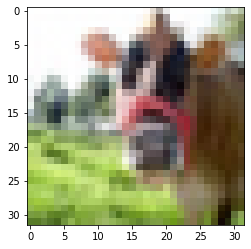}
     }
     \hfill
     \subfloat[ori.\label{cifar100-onelayer5}]{%
    \includegraphics[width=0.09\textwidth]{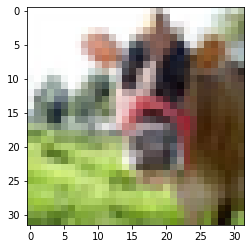}
     }
     \hfill
\subfloat[L1 distance]{
\begin{tabular}{lll}
\multicolumn{1}{c}{\bf Filters}  &\multicolumn{1}{c}{\bf Params. ($(b^{\prime})^2h$)} &\multicolumn{1}{c}{\bf Mean L1 error.}\\ 
\centering
1  &64 &2.4\\
5 & 1280&0.21\\
11&2816&0.04\\
12 &3072 &0.00019\\
\end{tabular}\label{table1}
    }
    \caption{\textit{One-layer CNN}: we show that the image reconstruction changes with increasing number of filters (numerical result is shown in Table \ref{table1}).}
    \label{1layer_loss_params}
    \vskip -2mm
\end{figure}

\begin{figure}[!ht]
    \centering
    \subfloat[Rec. using \ac{mlp} without regularizer\label{cifar100-bth16-mlp1}]{%
    \includegraphics[width=0.45\textwidth]{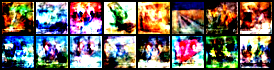}
     }
    \subfloat[Rec. using  \ac{mlp} with orthogonality regularizer\label{cifar100-bth16-mlp2}]{%
    \includegraphics[width=0.45\textwidth]{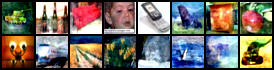}
     } 
     
    \subfloat[original input\label{cifar100-bth16-cnn1}]{%
    \includegraphics[width=0.45\textwidth]{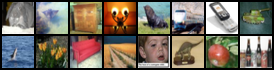}
     }
    \subfloat[Rec. using \ac{cnn} with L2 regularizer\label{cifar100-bth16-cnn2}]{%
    \includegraphics[width=0.45\textwidth]{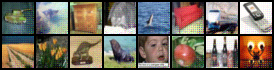}
     }  
    \caption{Batch size 16 (\ac{mlp} and \ac{cnn}).}
    \label{cifar100-bth16-reg}
    \vskip -4mm 
\end{figure}

\begin{figure*}[!ht]
    \centering
    \subfloat[Reconstruction with L2 regularizer\label{fig:cifar100-mlp-bth100-reg}]{%
    \includegraphics[width=0.3\textwidth]{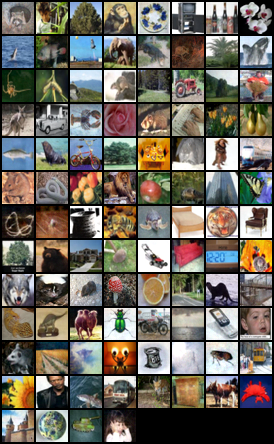}
     }
     \hfill
    \subfloat[Reconstruction without regularizer\label{fig:cifar100-mlp-bth100-no-reg}]{%
    \includegraphics[width=0.3\textwidth]{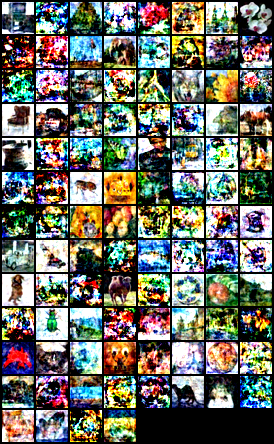}
     }
      \hfill
    \subfloat[Original input]{%
    \includegraphics[width=0.3\textwidth]{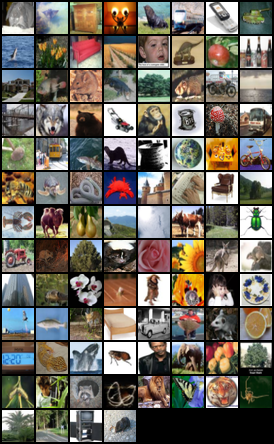}
     }
     \vskip -2mm
    \caption{\ac{mlp} reconstruction with batch size 100 (CIFAR100).}
    \label{fig:cifar100-mlp-bth100}
\end{figure*}

\textbf{\ac{mlp} Reconstruction Analysis.} We implement one-instance reconstruction on \ac{mlp} by the analytical form $x_i=\frac{\partial \ell}{\partial w_{1i}^1}/\frac{\partial \ell}{\partial b_1^1}, \forall i\in[1,d]$, using one-hidden layer \ac{mlp} with only one unit in hidden layer. The outcome is demonstrated in Figure \ref{imagenet-single} using ImageNet and \ac{wbc} dataset, the average L1 distance per pixel between original input and reconstruction is bounded below $1\mathrm{e}-8$, which is almost lossless with $\mathcal{O}(n)$ complexity. For \ac{mlp} batch reconstruction, we experimentally set the batch size equal to four, identical to the number of units in hidden layer, as shown in Figure \ref{fig:regularizer}. We first give the final reconstruction of four inputs without regularizer in Figure \ref{mnist:noregularizer}, whereas in Figure \ref{mnist:regularizer} the reconstruction quality is being improved significantly with the orthogonality regularizer. More specifically, we start $\lambda$ from $0.1$ and gradually decay after 200 epochs by $90\%$. Besides, from Figure \ref{mnist:regularizer-1} to \ref{mnist:regularizer-4}, we partially show the reconstruction procedure and we can see how the regularizer plays the key role during optimization procedure to hinder the similarities between instances. Moreover, we also show \ac{mlp} reconstruction with batch size 16 in Figure \ref{cifar100-bth16-mlp1} and \ref{cifar100-bth16-mlp2}. Empirically, it shows that the optimization for high-dimension demixing is very challenging and the existence of regularizer significantly improves the numerical stability and the reconstruction quality. In Figure~\ref{fig:cifar100-mlp-bth100}, we demonstrate that Proposition~\ref{batch-mlp} is also valid when batch size $B$ is large, eg., $B=100$.

\textbf{\ac{cnn} Reconstruction Analysis.} We divide one-instance \ac{cnn} reconstruction into two steps: 1) we directly compute the output of convolutional layer; 2) apply iterative optimization. We test the reconstruction performance with different numbers of kernels. We let the kernel size be 5, padding size 2, stride size 2 and pooling size 1, thus at least 12 kernels are required according to Proposition \ref{cnn-theorem}. In Figure \ref{1layer_loss_params} we first visually show the reconstruction improvement with the incremental numbers of kernels, and from Table \ref{table1} we numerically show reconstruction error (L1 distance) decreasing with more kernels. For batch \ac{cnn} reconstruction, we demonstrate it in Figure \ref{cifar100-bth16-cnn1} and \ref{cifar100-bth16-cnn2} where we have the similar convolutional setup with one-instance \ac{cnn}, thus 12 kernels are applied in convolutional layer. Note for the redundant architectures, i.e., with more kernels and units than required, we can easily mask the corresponding gradients during iterative optimization to fasten the reconstruction step. 

\vskip -5mm
\section{Conclusion}\label{sec:conclusion}
\vskip -4mm
We theoretically studied the minimal structural requirements for reconstruction and analyzed the relations between network architecture, size, and reconstruction quality. We show that the number of units in first hidden layer should be equal to greater than batch size using \ac{mlp}. It is worthwhile to mention that joint training using \ac{mlp} (with bias) with batch size equal to one in \ac{fl} is extremely vulnerable to the adversary. For \ac{cnn}, the number of kernels along with the number of units in a fully-connected (dense) layer decides the quality of reconstruction. Specifically, number of units in hidden layer is determined by batch size and required number of kernels is decided by output dimension after convolutional layers. Our observations also apply to big batch size with the aid of the regularizer. We hope that the limits explored in the present work and conditions for reconstruction can aid the practitioners to choose network architecture and communication strategies when applying \ac{fl} on sensitive information-related applications, e.g., medical diagnosis, stocking price prediction.

\section{Acknowledgement}
The research leading to these results has received funding from the European Union’s Horizon 2020 research and innovation program under the Marie Sklodowska-Curie grant agreement No. 764785, FORA-Fog Computing for Robotics and Industrial Automation.

\bibliography{sample}

\newpage
\section{Appendix A}\label{one-mlp-appen}
\begin{proposition}[\textsc{one-instance \ac{mlp} reconstruction}]

To reconstruct one input based on \ac{mlp}, we derive the analytical form to compute the (almost) lossless input, with only \textbf{single} unit in the first hidden layer as long as bias term exists, regardless how deep the network is.
\end{proposition}
\vskip -8mm
\begin{proof}\label{proof-mlp1}
\vskip -3mm
The derivative of loss function $\ell$ w.r.t $a_j$ (output of network $f_w$) is $p_j-y_j$ by plugging eq. (\ref{softmax_derive}) into eq. (\ref{loss_derive}).
\begin{equation}\label{softmax_derive}
\frac{\partial p_i}{\partial a_j} = \left\{
    \begin{array}{ll}
         p_j(1-p_j), \quad i=j\\
    -p_j p_i, \quad i\neq j
    \end{array}
\right.
\end{equation}
\vskip -3mm
\begin{equation}\label{loss_derive}
 \begin{split}
 \frac{\partial \ell}{\partial a_j}&=-\sum_k y_k\frac{\partial \log p_k}{\partial a_j}= -\sum_k y_k \frac{1}{p_k}\frac{\partial p_k}{\partial a_j}\\
 &=-[\frac{y_j}{p_j}(p_j(1-p_j))-\sum_{k\neq j}\frac{y_k}{p_k}p_j p_k] \\
  &= p_j-y_j
 \end{split}
\end{equation} 
Thus, all the partial derivatives of Jacobian matrix are as the followings:
\begin{equation}\label{bias_output_layer}
    \frac{\partial \ell}{\partial b_j^2} =p_{.j}-y_{.j} \quad \forall j\in[1,n_2]
\end{equation}
\vskip -4mm
\begin{equation}\label{weight_output_layer}
\begin{split}
      \frac{\partial \ell}{\partial w_{ji}^2} &= (p_{.j}-y_{.j})\sigma(w^1_ix+b^1_i)= (p_{.j}-y_{.j})\sigma_i \\ 
      & \forall j\in [1,n_2], \forall i \in [1,n_1] 
\end{split}
\end{equation}
\vskip -4mm
\begin{equation}\label{bias_hidden_layer}
    \frac{\partial \ell}{\partial b_j^1}=\sum_i^{n_2}(p_{.i}-y_{.i})w^2_{ij}\sigma^{\prime}(w^1_jx+b^1_j)=\sum_i^{n_2}(p_{.i}-y_{.i})w^2_{ij}\sigma^{\prime}_j
\end{equation}
\vskip -4mm
\begin{equation}\label{weight_hidden_layer}
    \frac{\partial \ell}{\partial w_{ji}^1}=\sum_{k}^{n_2}(p_{.k}-y_{.k})w^2_{kj}\sigma^{\prime}_jx_i
\end{equation}
We can directly compute $x_i=\frac{\partial \ell}{\partial w_{1i}^1}/\frac{\partial \ell}{\partial b_1^1}, \forall i\in[1,d]$ from eq. (\ref{bias_hidden_layer}) and (\ref{weight_hidden_layer}) where $j$ is equal to one since only one unit is required in hidden layer.  
\end{proof}

To generalize it to a model with $L$ layers, we only need the derivatives w.r.t the weights and bias in the first hidden layer. They are expressed as:
\begin{equation}
\begin{split}
    \frac{\partial \ell}{\partial w^1_n}&=\sum_{h}^{n_L}(p_{.h}-y_{.h})\sum_{j}^{n_{L-1}}w_{hj}^{L}(\sigma_j^{L-1})^{\prime}\cdots w^2_{mn}(\sigma^1_n)^{\prime}x\\
    \frac{\partial \ell}{\partial b^1_n}&=\sum_{h}^{n_{L}}(p_{.h}-y_{.h})\sum_{j}^{n_{L-1}}w_{hj}^{L}(\sigma_j^{L-1})^{\prime}\cdots w^2_{mn}(\sigma^1_n)^{\prime}
\end{split}
\end{equation}
where $n_L$ is the number of nodes in layer $L$. Then, the reconstruction is the division between them.

\section{Appendix B}\label{proof-cnn}
For a single convolutional layer \ac{cnn}, the kernel parameters (kernel size $k$, padding size $p$, stride size $s$) determine the output size $d^{\prime}$ after convolutional layers. Say we have input $X\in \rbb^{B\times C\times d\times d} $, for the simplicity we assume height and width are identical, and $C$ is the channel number and $B$ is the batch size. We define a square kernel with width $k$ (weights indicated as $l^0$), bias term $r$, and we have $h$ kernels. After convolutional layer, the width of output is $d^{\prime}=\frac{d+2p-k}{s}+1$.


Then we define $H = [\text{vec}(\hat{z}_{\underline{1}..}),\text{vec}(\hat{z}_{\underline{2}..}),...,\text{vec}(\hat{z}_{\underline{h}..})]$, where $\hat{z}_i$ is the output of convolutional layer, and $|H|=h(d^{\prime})^2=n_0$, after convolutional layer we have one hidden layer and an output layer. It is therefore expressed as $p_j = \sum_{i=1}^{n_1}w_{ji}^2\sigma(w^1_i H+b^1_i)+b^2_j$, where $w^1\in \rbb^{n_1\times n_0}$, $w^2\in \rbb^{n_2\times n_1}$, $b^1\in \rbb^{n_1}$ and $b^2\in \rbb^{n_2}$ are the weights and bias in the hidden and output layer, $\sigma()$ is sigmoid function as defined before. Thus, we have the derivatives w.r.t $w^1,b^1,w^2,b^2$ shown in eq. (\ref{bias_output_layer}), (\ref{weight_output_layer}), (\ref{bias_hidden_layer}) and (\ref{weight_hidden_layer}). Moreover, the partial derivatives w.r.t $r$ and $l^0$ are shown in eq. (\ref{cnn_bias_conv_layer}) and (\ref{cnn_weight_conv_layer}).
\begin{equation}\label{cnn_bias_conv_layer}
    \frac{\partial \ell}{\partial r_{\underline{m}}}=\sum_{i=1}^{d^{\prime}}\sum_{j=1}^{d^{\prime}}\frac{\partial \ell}{\partial z_{\underline{m}ij}} \frac{\partial z_{\underline{m}ij}}{\partial b_{\underline{m}}^1} =\sum_{i=1}^{d^{\prime}}\sum_{j=1}^{d^{\prime}} \frac{\partial \ell}{\partial z_{\underline{m}ij}}\quad \forall \underline{m}\in[1,h]
\end{equation}
\vskip -5mm
\begin{equation}\label{cnn_weight_conv_layer}
    \frac{\partial \ell}{\partial l_{\underline{m}cgh}^0}=\sum_{i=1}^{d^{\prime}} \sum_{j=1}^{d^{\prime}}\frac{\partial \ell}{\partial z_{\underline{m}ij}}\hat{x}_{c,si+g-1,sj+h-1} \quad \forall \underline{m}\in[1,h]
\end{equation}
\vskip -4mm
\begin{equation}
 \frac{\partial \ell}{\partial \hat{z}_{\underline{m}ij}}=\frac{\partial \ell}{\partial H[(m-1)\times (d^{\prime})^2+(i-1)\times d^{\prime}+j]}   
\end{equation}

Generalizing it to the multiple-convolutional-layer neural network, the output $H^{i-1}$ of layer $i-1$ is the input of layer $i$, then we can express it as the following, where $h^{i-1}$ and $k^{i-1}$ are the number of kernels and kernel width in layer $i-1$.
\begin{equation}
H^{i}_{mij} = (\sum_{c=1}^{h^{i-1}}\sum_{g=1}^{k^{i-1}}\sum_{n=1}^{k^{i-1}}l^i_{mcgn}H^{i-1}_{c,si+g-1,sj+n-1})+b^{i}_m    
\end{equation}
We can recursively calculate the number of kernels required in each convolutional layer from right to left order (from the last convolutional layer to the first one).

\newpage
\section{Appendix C}
\begin{figure}[ht] 
\centering
\begin{subfigure}{0.1\textwidth}
\includegraphics[width=\linewidth]{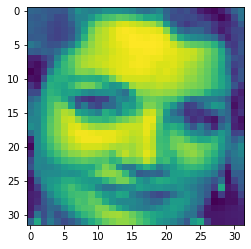}
\end{subfigure}\hspace*{\fill}
\begin{subfigure}{0.1\textwidth}
\includegraphics[width=\linewidth]{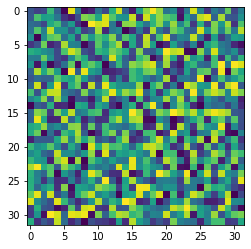}
\end{subfigure}\hspace*{\fill}
\begin{subfigure}{0.1\textwidth}
\includegraphics[width=\linewidth]{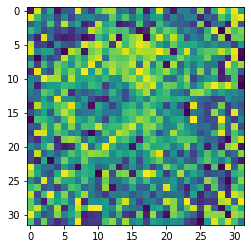}
\end{subfigure}\hspace*{\fill}
\begin{subfigure}{0.1\textwidth}
\includegraphics[width=\linewidth]{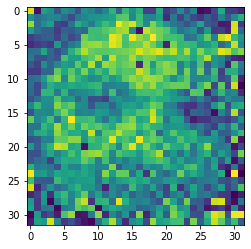}
\end{subfigure}\hspace*{\fill}
\begin{subfigure}{0.1\textwidth}
\includegraphics[width=\linewidth]{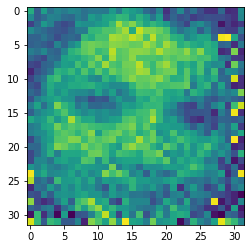}
\end{subfigure}\hspace*{\fill}
\begin{subfigure}{0.1\textwidth}
\includegraphics[width=\linewidth]{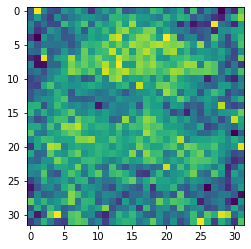}
\end{subfigure}\hspace*{\fill}
\begin{subfigure}{0.1\textwidth}
\includegraphics[width=\linewidth]{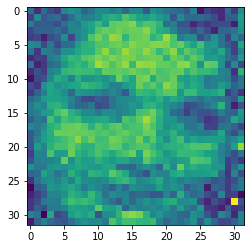}
\end{subfigure}\hspace*{\fill}
\begin{subfigure}{0.1\textwidth}
\includegraphics[width=\linewidth]{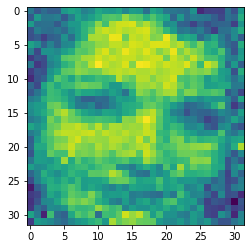}
\end{subfigure}\hspace*{\fill}
\begin{subfigure}{0.1\textwidth}
\includegraphics[width=\linewidth]{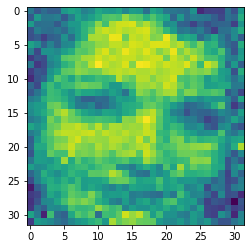}
\end{subfigure}\hspace*{\fill}

\begin{subfigure}{0.1\textwidth}
\includegraphics[width=\linewidth]{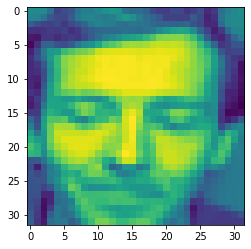}
\end{subfigure}\hspace*{\fill}
\begin{subfigure}{0.1\textwidth}
\includegraphics[width=\linewidth]{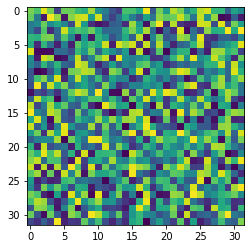}
\end{subfigure}\hspace*{\fill}
\begin{subfigure}{0.1\textwidth}
\includegraphics[width=\linewidth]{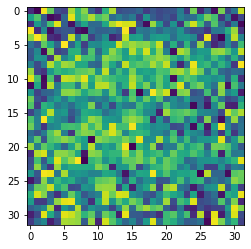}
\end{subfigure}\hspace*{\fill}
\begin{subfigure}{0.1\textwidth}
\includegraphics[width=\linewidth]{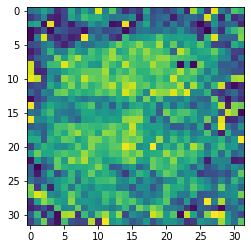}
\end{subfigure}\hspace*{\fill}
\begin{subfigure}{0.1\textwidth}
\includegraphics[width=\linewidth]{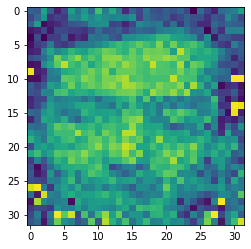}
\end{subfigure}\hspace*{\fill}
\begin{subfigure}{0.1\textwidth}
\includegraphics[width=\linewidth]{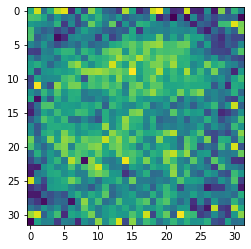}
\end{subfigure}\hspace*{\fill}
\begin{subfigure}{0.1\textwidth}
\includegraphics[width=\linewidth]{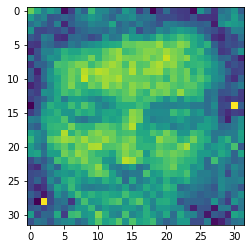}
\end{subfigure}\hspace*{\fill}
\begin{subfigure}{0.1\textwidth}
\includegraphics[width=\linewidth]{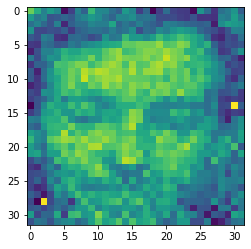}
\end{subfigure}\hspace*{\fill}
\begin{subfigure}{0.1\textwidth}
\includegraphics[width=\linewidth]{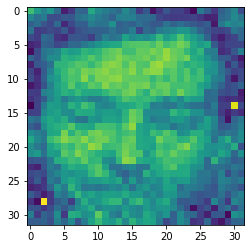}
\end{subfigure}\hspace*{\fill}

\begin{subfigure}{0.1\textwidth}
\includegraphics[width=\linewidth]{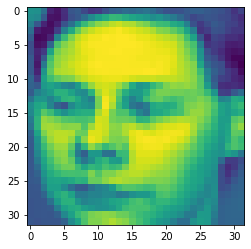}
\end{subfigure}\hspace*{\fill}
\begin{subfigure}{0.1\textwidth}
\includegraphics[width=\linewidth]{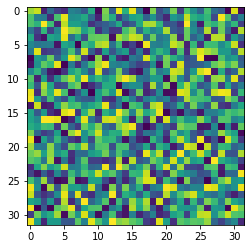}
\end{subfigure}\hspace*{\fill}
\begin{subfigure}{0.1\textwidth}
\includegraphics[width=\linewidth]{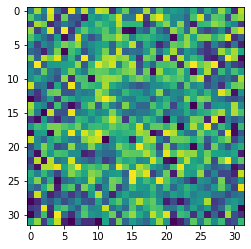}
\end{subfigure}\hspace*{\fill}
\begin{subfigure}{0.1\textwidth}
\includegraphics[width=\linewidth]{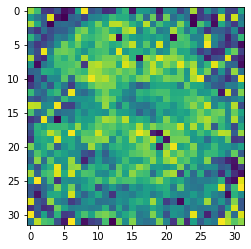}
\end{subfigure}\hspace*{\fill}
\begin{subfigure}{0.1\textwidth}
\includegraphics[width=\linewidth]{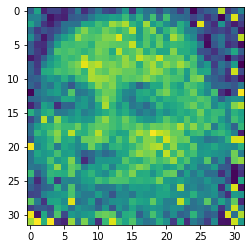}
\end{subfigure}\hspace*{\fill}
\begin{subfigure}{0.1\textwidth}
\includegraphics[width=\linewidth]{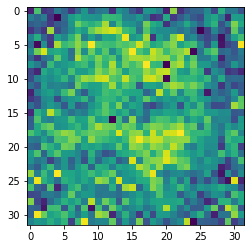}
\end{subfigure}\hspace*{\fill}
\begin{subfigure}{0.1\textwidth}
\includegraphics[width=\linewidth]{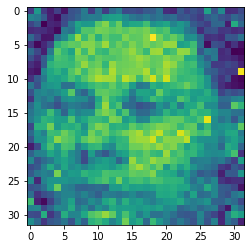}
\end{subfigure}\hspace*{\fill}
\begin{subfigure}{0.1\textwidth}
\includegraphics[width=\linewidth]{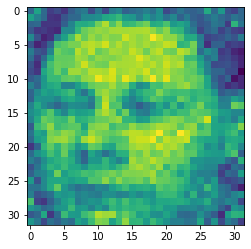}
\end{subfigure}\hspace*{\fill}
\begin{subfigure}{0.1\textwidth}
\includegraphics[width=\linewidth]{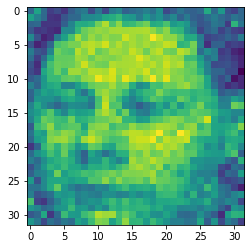}
\end{subfigure}\hspace*{\fill}

\begin{subfigure}{0.1\textwidth}
\includegraphics[width=\linewidth]{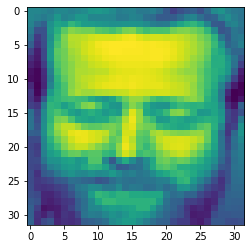}
\end{subfigure}\hspace*{\fill}
\begin{subfigure}{0.1\textwidth}
\includegraphics[width=\linewidth]{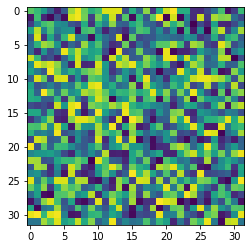}
\end{subfigure}\hspace*{\fill}
\begin{subfigure}{0.1\textwidth}
\includegraphics[width=\linewidth]{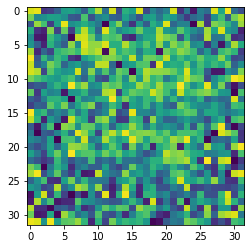}
\end{subfigure}\hspace*{\fill}
\begin{subfigure}{0.1\textwidth}
\includegraphics[width=\linewidth]{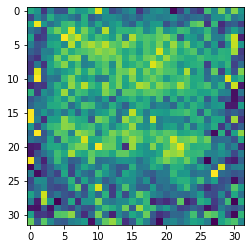}
\end{subfigure}\hspace*{\fill}
\begin{subfigure}{0.1\textwidth}
\includegraphics[width=\linewidth]{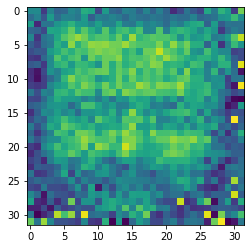}
\end{subfigure}\hspace*{\fill}
\begin{subfigure}{0.1\textwidth}
\includegraphics[width=\linewidth]{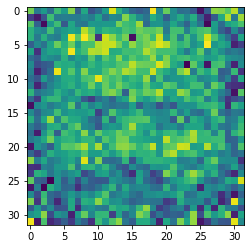}
\end{subfigure}\hspace*{\fill}
\begin{subfigure}{0.1\textwidth}
\includegraphics[width=\linewidth]{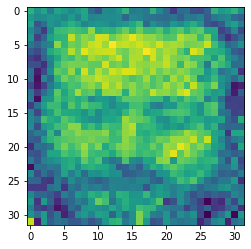}
\end{subfigure}\hspace*{\fill}
\begin{subfigure}{0.1\textwidth}
\includegraphics[width=\linewidth]{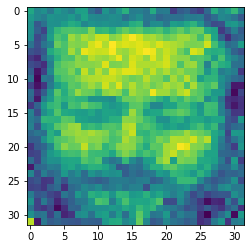}
\end{subfigure}\hspace*{\fill}
\begin{subfigure}{0.1\textwidth}
\includegraphics[width=\linewidth]{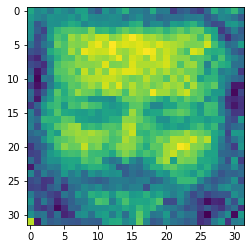}
\end{subfigure}\hspace*{\fill}

\begin{subfigure}{0.1\textwidth}
\includegraphics[width=\linewidth]{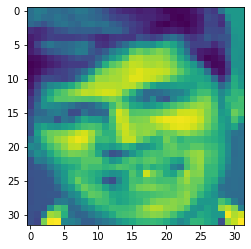}
\caption{ori.}
\end{subfigure}\hspace*{\fill}
\begin{subfigure}{0.1\textwidth}
\includegraphics[width=\linewidth]{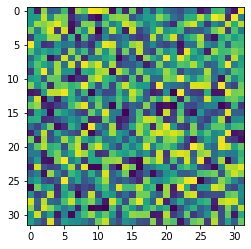}
\caption{10}
\end{subfigure}\hspace*{\fill}
\begin{subfigure}{0.1\textwidth}
\includegraphics[width=\linewidth]{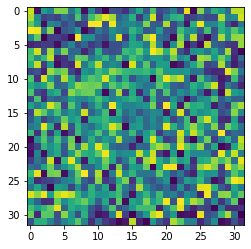}
\caption{50}
\end{subfigure}\hspace*{\fill}
\begin{subfigure}{0.1\textwidth}
\includegraphics[width=\linewidth]{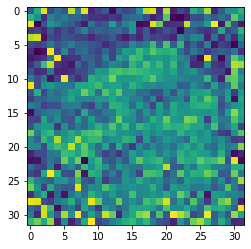}
\caption{90}
\end{subfigure}\hspace*{\fill}
\begin{subfigure}{0.1\textwidth}
\includegraphics[width=\linewidth]{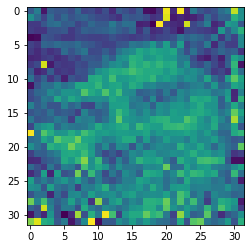}
\caption{final}
\end{subfigure}\hspace*{\fill}
\begin{subfigure}{0.1\textwidth}
\includegraphics[width=\linewidth]{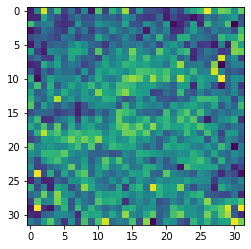}
\caption{10(*)}
\end{subfigure}\hspace*{\fill}
\begin{subfigure}{0.1\textwidth}
\includegraphics[width=\linewidth]{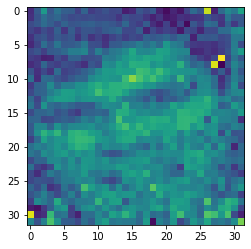}
\caption{50(*)}
\end{subfigure}\hspace*{\fill}
\begin{subfigure}{0.1\textwidth}
\includegraphics[width=\linewidth]{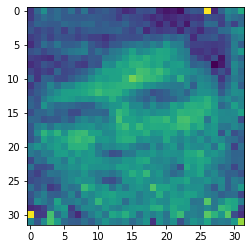}
\caption{90(*)}
\end{subfigure}\hspace*{\fill}
\begin{subfigure}{0.1\textwidth}
\includegraphics[width=\linewidth]{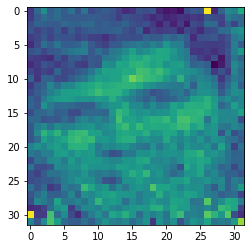}
\caption{fi.(*)}
\end{subfigure}\hspace*{\fill}

\caption{\textit{Face} (batch reconstruction):  Mini-batch contains 5 images, and we show partial reconstruction for 10, 50, 90, and final (400 iterations, with L-BFGS optimizer)iteration for \cite{zhu2019deep} and our method accordingly.}
\label{face_prior}
\end{figure}

\begin{figure}[ht]
     \subfloat[Ori.\label{subfig1:fmri}]{%
      \includegraphics[width=0.15\textwidth]{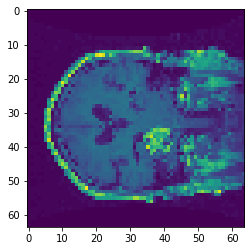}
     }
     \hfill
     \subfloat[20 itrs\label{subfig2:fmri}]{%
      \includegraphics[width=0.15\textwidth]{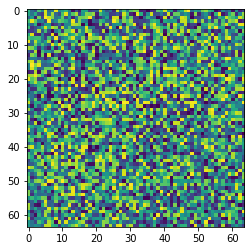}
     }
     \hfill
      \subfloat[80 itrs\label{subfig3:fmri}]{%
      \includegraphics[width=0.15\textwidth]{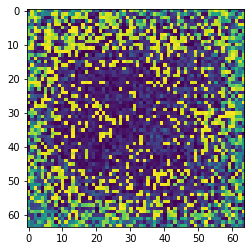}
     }
     \hfill
     \subfloat[final\label{subfig4:fmri}]{%
      \includegraphics[width=0.15\textwidth]{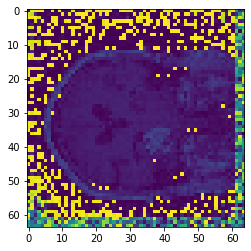}
     }
     
      \subfloat[ori.\label{subfig5:fmri}]{%
      \includegraphics[width=0.15\textwidth]{images/fmri-ori1.png}
     }
     \hfill
    \subfloat[20 itrs(our)\label{subfig6:fmri}]{%
      \includegraphics[width=0.15\textwidth]{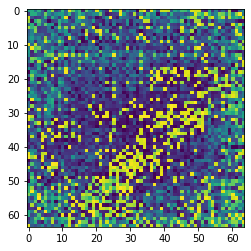}
     }
     \hfill
      \subfloat[80 itrs(our)\label{subfig7:fmri}]{%
      \includegraphics[width=0.15\textwidth]{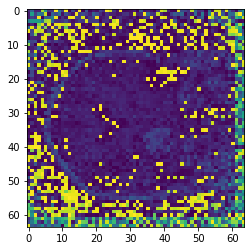}
     }
     \hfill
      \subfloat[final(our)\label{subfig8:fmri}]{%
      \includegraphics[width=0.15\textwidth]{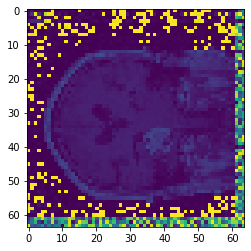}
     }
     \caption{\textit{fMRI}: The plots (first row) are produced by \cite{zhu2019deep}, and it shows the reconstructions after 20, 80, and final iteration accordingly, whereas the second row is our reconstruction.}
     \label{fmri_prior}
  \end{figure}

\begin{figure}[!ht]
    \centering
    \subfloat[fMRI reconstruction error\label{fmri-l1-com}]{%
    \includegraphics[width=0.39\textwidth]{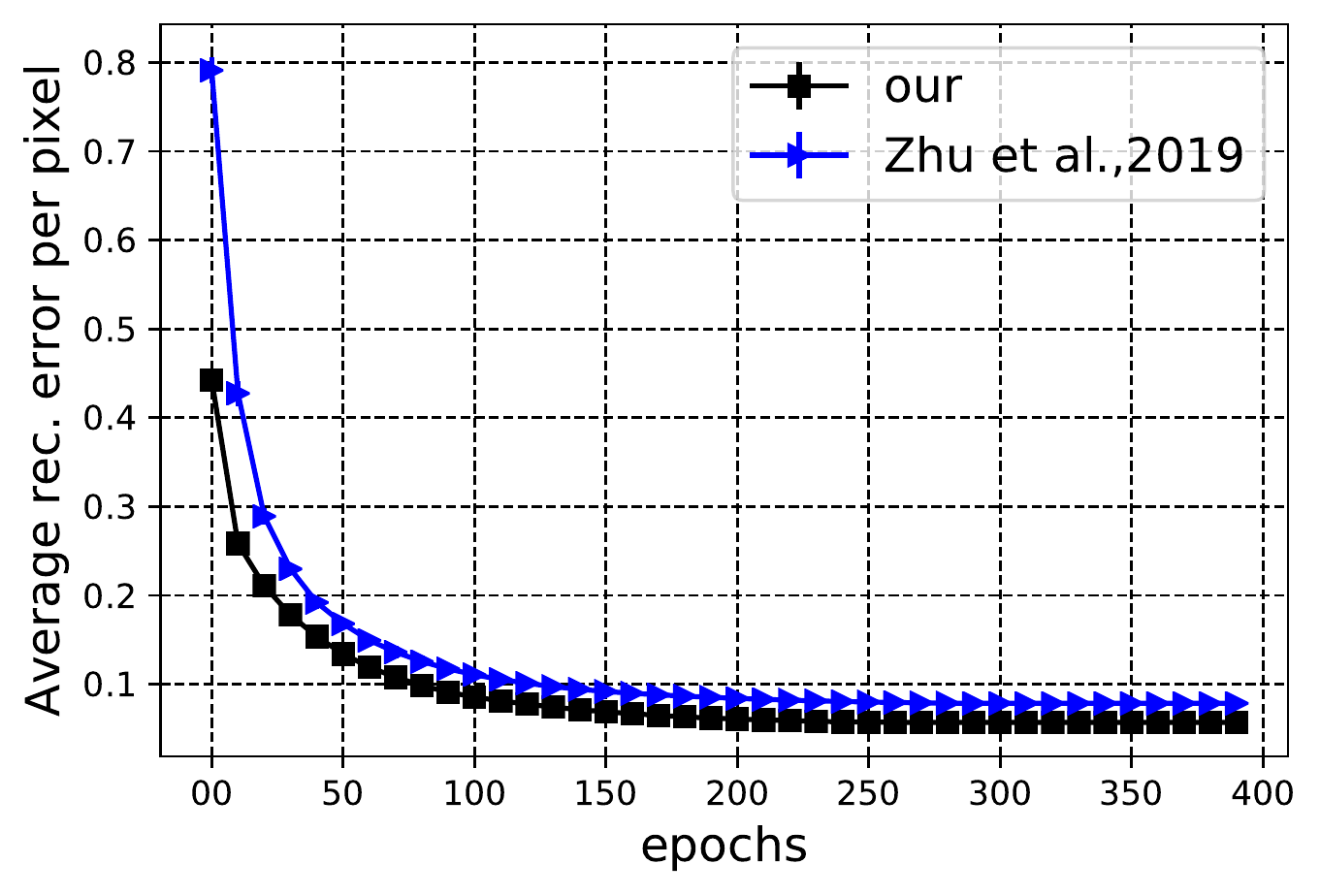}
     }
    \subfloat[Face reconstruction error\label{face-l1-com}]{%
    \includegraphics[width=0.4\textwidth]{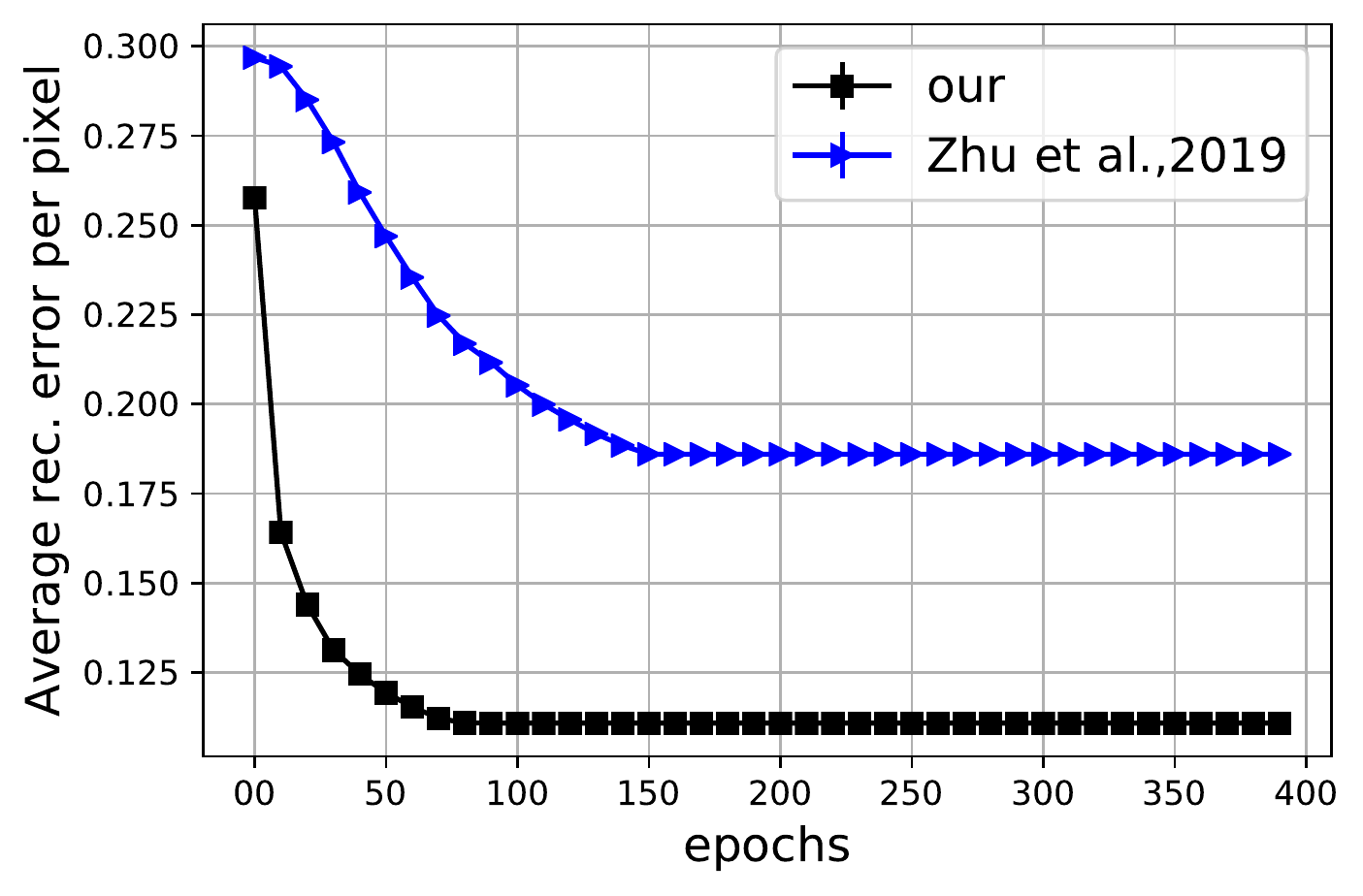}
     } 
    \caption{Trajectory of average error per pixel over the epoch number. This is the numerical comparison of Figure~\ref{face_prior} and \ref{fmri_prior}. For the fair comparison, we also use L-BFGS optimizer as~\citep{zhu2019deep}.}
    \label{rec-loss-com}
\end{figure}

fMRI image (available \href{https://figshare.com/articles/dataset/brain_tumor_dataset/1512427}{here}) set is the brain tumor dataset containing 3064 images from 233 patients with three types of brain tumor (meningioma, glioma, pituitary). Face dataset contains 40 individuals, and every image has size $1\times32\times32$.
We plot the snapshots of the reconstruction procedure in Figure~\ref{face_prior} and \ref{fmri_prior} for the visual comparison. The corresponding numerical results are demonstrated in Figure~\ref{rec-loss-com}.

\begin{figure}[ht]
\centering
\begin{minipage}{.9\textwidth}
 \subfloat[MNIST\label{mnist-reg}]{%
    \includegraphics[width=0.49\textwidth]{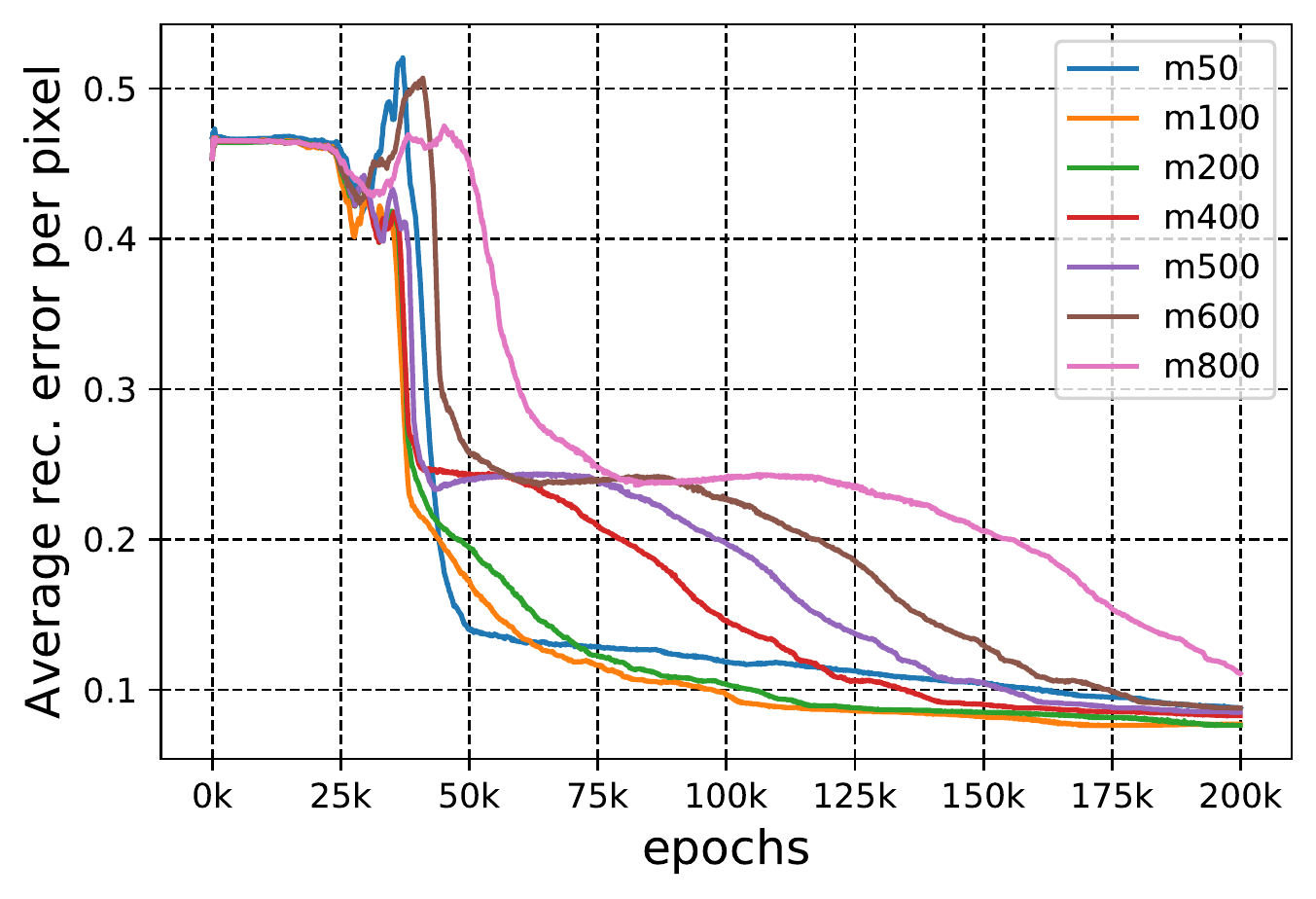}
     }
     \hfill
 \subfloat[KMNIST\label{kmnist-reg}]{%
    \includegraphics[width=0.44\textwidth]{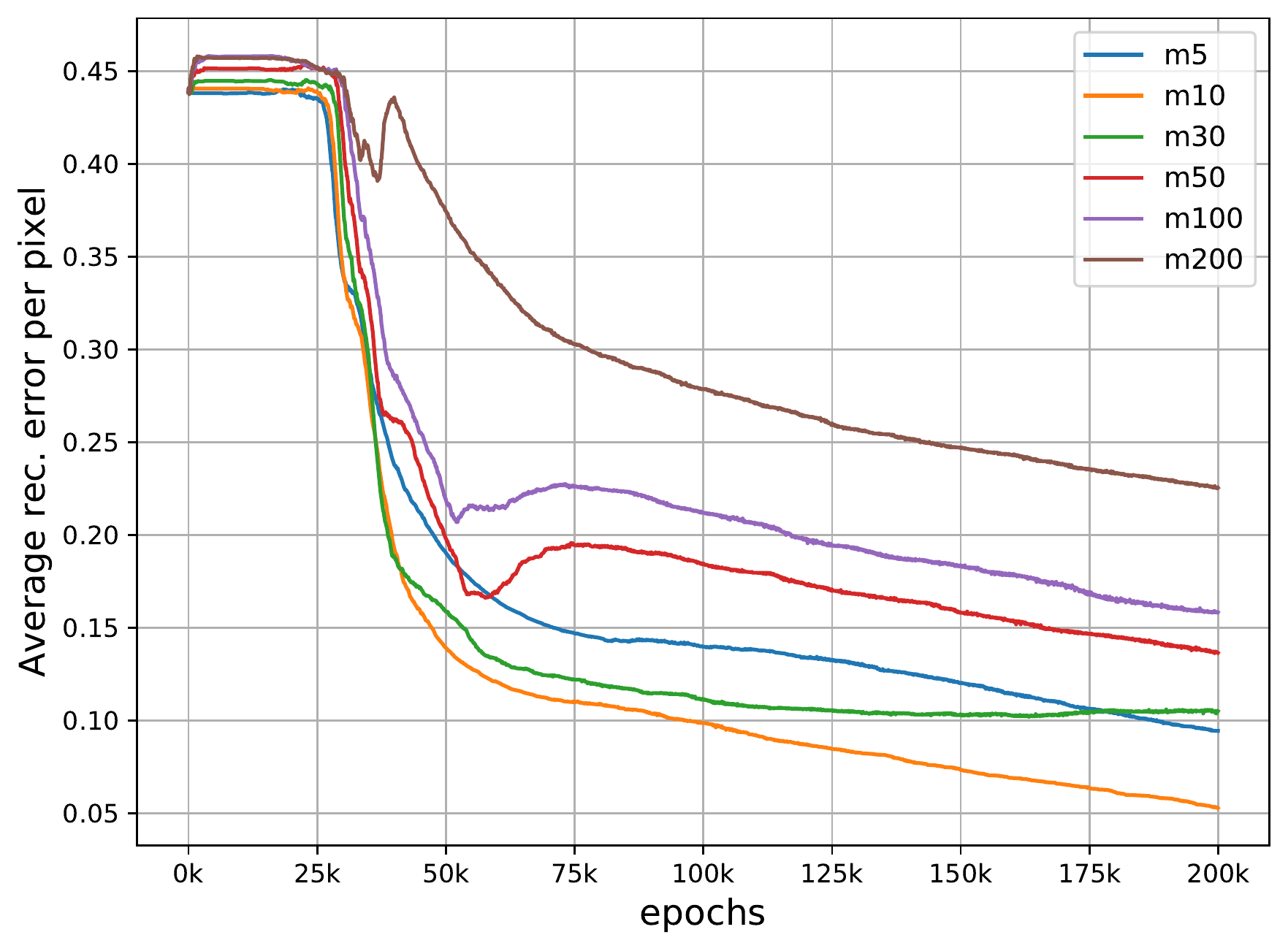}
     }
\caption{ Orthogonality regularizer with different interval values to decay.}
 \label{regularizer}
\end{minipage}%
\end{figure}

Sometimes the choice of $m$ (interval that orthogonality regularizer $\lambda$ decays) is crucial, for instance in Figure \ref{kmnist-reg} the reconstruction performance is significantly distinct with different values. While, in Figure \ref{mnist-reg}, the choice of $m$ is insensitive. 

\begin{figure}[ht]
    \centering
\begin{minipage}{.9\textwidth}
\subfloat[CIFAR100\label{cnn-com-cifar100}]{%
       \includegraphics[width=0.45\textwidth]{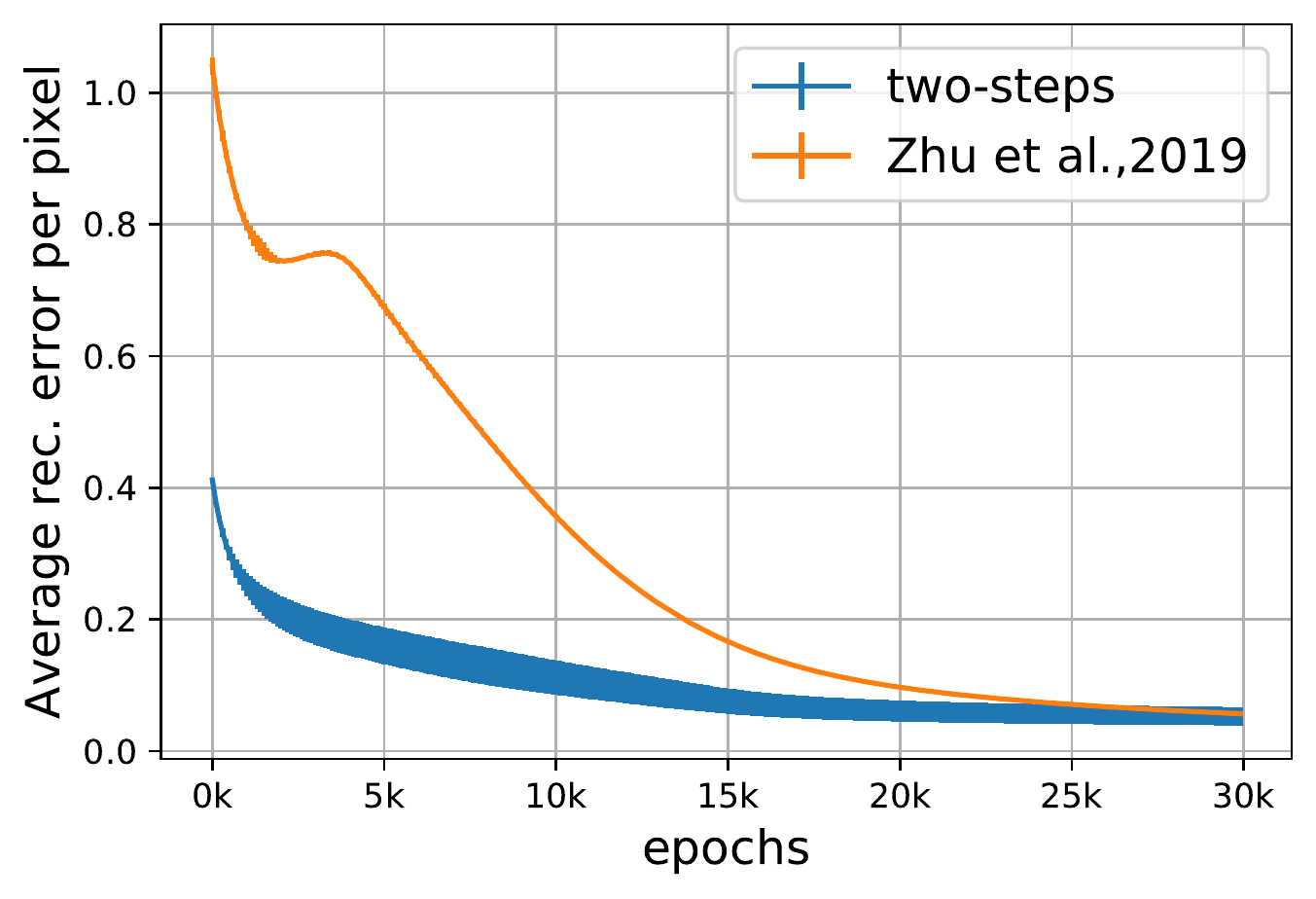}
     }
 \hfill
 \subfloat[MNIST\label{cnn-com-mnist}]{%
\includegraphics[width=0.45\textwidth]{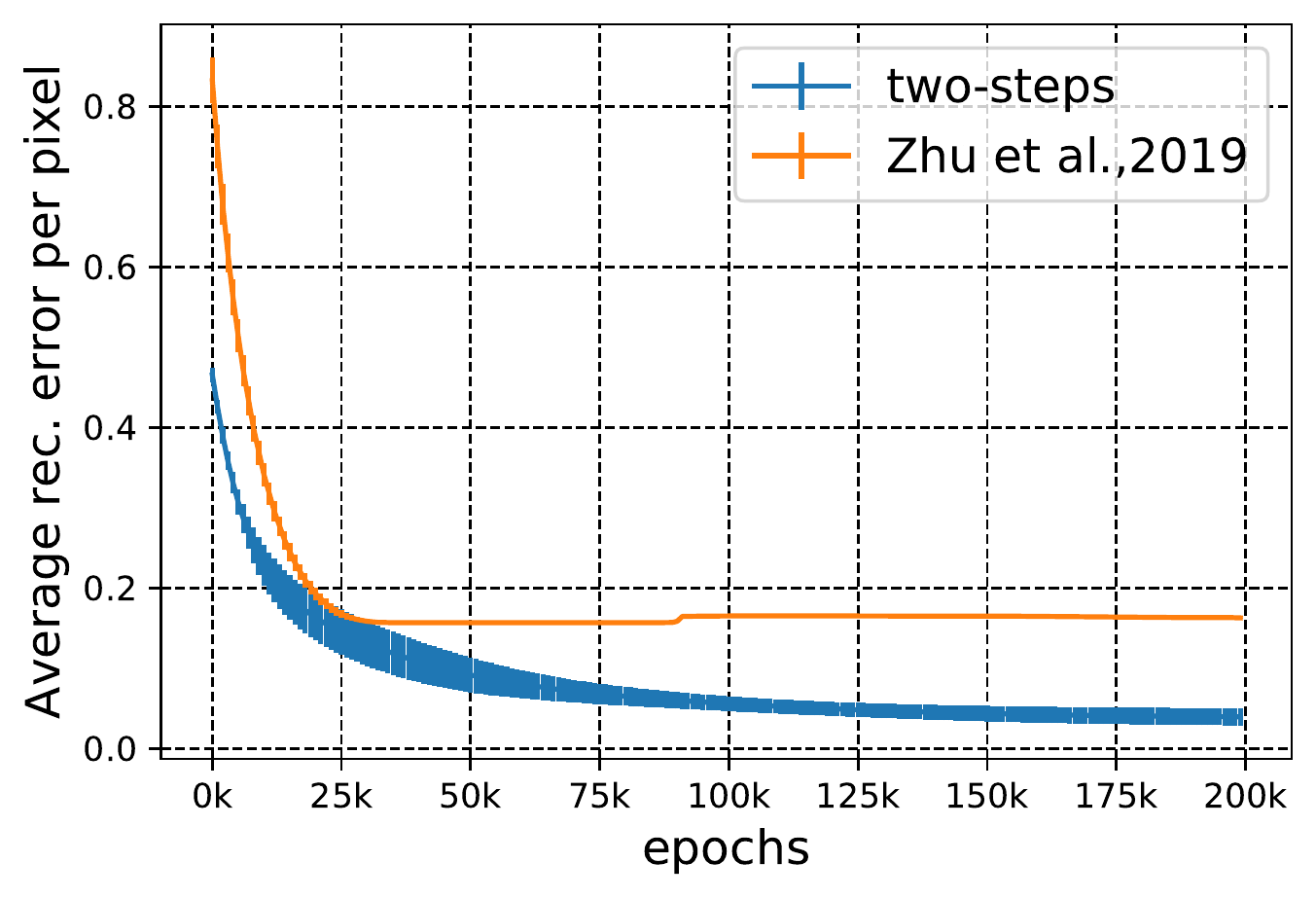}
     }
\caption{\ac{cnn} two-step reconstruction.}
\label{cnn-twosteps-com}
\end{minipage}
\end{figure}

The experimental results in Figure \ref{cnn-twosteps-com} show that the two-step reconstruction enables a faster convergence and lower error compared with \cite{zhu2019deep}. 

\end{document}

%% file: 00Acron.tex
\acrodef{fcp}[FCP]{fog computing platform}
\acrodef{cc}[CC]{Cloud Computing}
\acrodef{fc}[FC]{Fog Computing}
\acrodef{ec}[EC]{Edge Computing}
\acrodef{eds}[EDs]{Edge Devices}
\acrodef{gdpr}[GDPR]{General Data Protection Regulation}

\acrodef{ai}[AI]{artificial intelligence}
\acrodef{iot}[IoT]{Internet of Things}

\acrodef{ml}[ML]{Machine Learning}
\acrodef{fl}[FL]{Federated Learning}
\acrodef{cnn}[CNN]{Convolutional Neural Network}
\acrodef{mlp}[MLP]{Multilayer Perceptron}
\acrodef{rnn}[RNN]{Recurrent Neural Network}
\acrodef{lstm}[LSTM]{Long short-term memory}

\acrodef{map}[MAP]{Maximum a Posterior}
\acrodef{sgd}[SGD]{Stochastic Gradient Descent}
\acrodef{gan}[GAN]{Generative Adversarial Network}

\acrodef{wbc}[WBC]{White Blood Cell}